\documentclass[manuscript,screen]{acmart}
\usepackage{makecell}
\usepackage{url}
 \usepackage[caption=false,font=footnotesize]{subfig}

\newcommand{\deepshovel}{DeepShovel}
\newcommand{\DDE}{DDE}
 \newcommand{\SubItem}[1]{
    {\setlength\itemindent{15pt} \item[-] #1}
 }

\AtBeginDocument{
  \providecommand\BibTeX{{
    \normalfont B\kern-0.5em{\scshape i\kern-0.25em b}\kern-0.8em\TeX}}}

 \setcopyright{acmcopyright}
\copyrightyear{2022}
\acmYear{2022}
\acmDOI{10.1145/1122445.1122456}

\begin{document}

\title{\deepshovel: An Online Collaborative Platform for Data Extraction in Geoscience Literature with AI Assistance}


\author{Shao Zhang}
\email{shaozhang@sjtu.edu.cn}
\affiliation{
  \institution{Shanghai Jiao Tong University}
  \city{Shanghai}
  \country{China}
}

\author{Yuting Jia}
\email{hnxxjyt@sjtu.edu.cn}
\affiliation{
  \institution{Shanghai Jiao Tong University}
  \city{Shanghai}
  \country{China}
}

\author{Hui Xu}
\email{xhui_1@sjtu.edu.cn}
\affiliation{
  \institution{Shanghai Jiao Tong University}
  \city{Shanghai}
  \country{China}
}

\author{Ying Wen}
\email{ying.wen@sjtu.edu.cn}
\affiliation{
  \institution{Shanghai Jiao Tong University}
  \city{Shanghai}
  \country{China}
}

\author{Dakuo Wang}
\email{dakuo.wang@ibm.com}
\affiliation{
  \institution{IBM Research}
  \city{Cambridge, Massachusetts}
  \country{United States}
}

\author{Xinbing Wang}
\email{xwang8@sjtu.edu.cn}
\affiliation{
  \institution{Shanghai Jiao Tong University}
  \city{Shanghai}
  \country{China}
}

\begin{teaserfigure}
    \includegraphics[width=\textwidth]{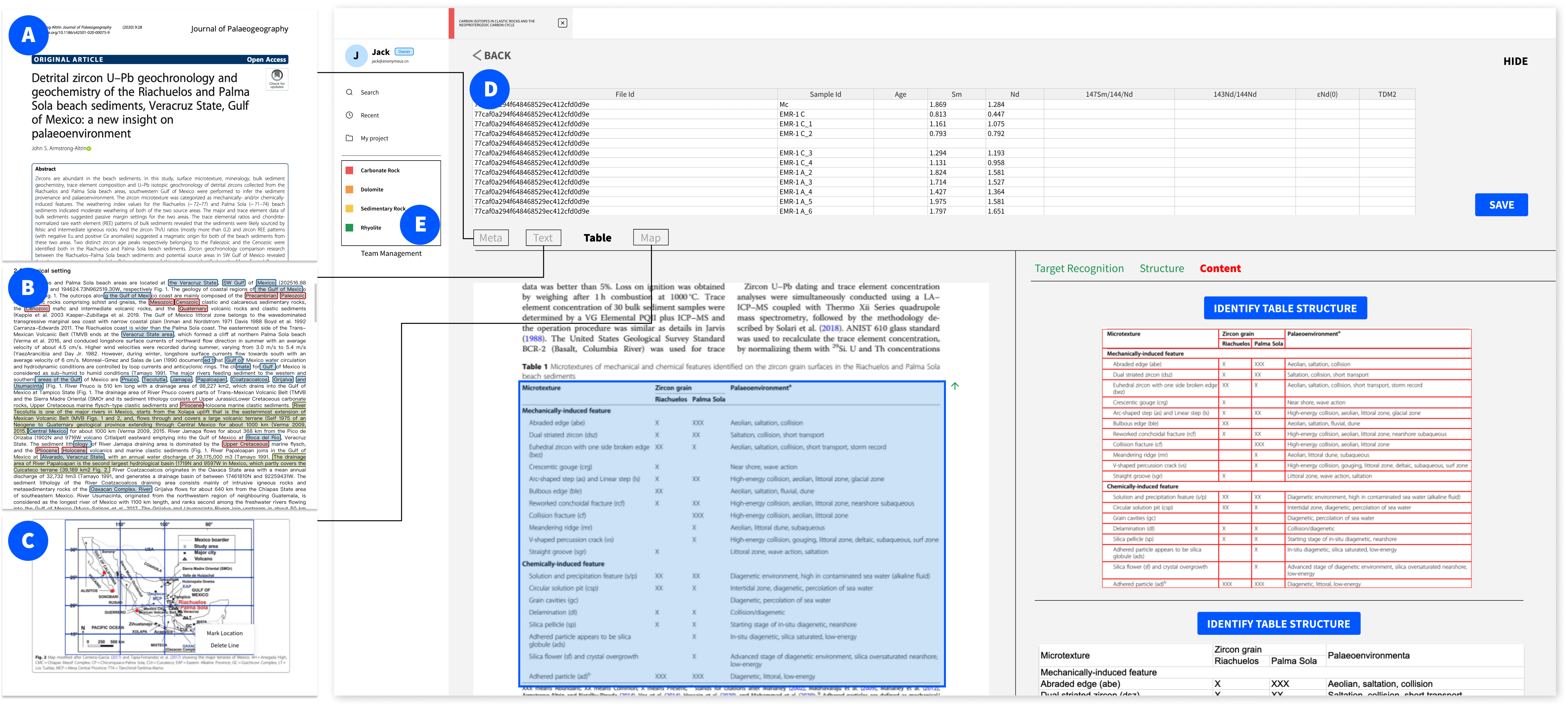}
    \caption{User Interface of \deepshovel, an online collaborative platform for data extraction in geoscience literature with AI assistance. The main part of this figure illustrates the table extraction and integration functions (D), while the system can also support meta information extraction (A), text extraction (B), map recognition and location extraction (C), and team and document management (E).}
    \Description{figure description}
    \label{fig:UI_overview}
\end{teaserfigure}

\renewcommand{\shortauthors}{Shao Zhang et al.}

\begin{abstract}
Geoscientists, as well as researchers in many fields, need to read a huge amount of literature to locate, extract, and aggregate relevant results and data to enable future research or to build a scientific database, but there is no existing system to support this use case well.
In this paper, based on the findings of a formative study about how geoscientists collaboratively annotate literature and extract and aggregate data, we proposed \deepshovel, a publicly-available AI-assisted data extraction system to support their needs. 
\deepshovel~ leverages the state-of-the-art neural network models to support researcher(s) easily and accurately annotate papers (in the PDF format) and extract data from tables, figures, maps, etc. in a human-AI collaboration manner. 
A follow-up user evaluation with 14 researchers suggested \deepshovel~ improved users' efficiency of data extraction for building scientific databases, and encouraged teams to form a larger scale but more tightly-coupled collaboration.

\end{abstract}


\maketitle

\section{Introduction}
The shifting of the data-driven research paradigm raises new requirements for researchers to build \textit{scientific databases} \cite{hoeppe2021encoding} in many disciplines, including Geoscience \cite{bergen2019machine}, Medicine \cite{austin2016application}, Biology \cite{altaf2014systems},  Chemistry \cite{schleder2019ab}.
Scientific database is a collection of structured and verified research results that consists of various numeric, word-oriented, or image-organized data, which plays a central role in data-driven research \cite{national2000question}.
In this paper, we focus on the need of constructing scientific databases in geoscience, which can help geoscientists to discover unknown phenomena and novel insights in Earth \cite{fan2020high,dirzo2014defaunation,tucker2018moving}.
To obtain enough high-quality data, geoscientists often review a large amount of published literature \cite{PUETZ2018877,mcmahon2018evolution,puetz2018statistical} (generally are PDF documents), from which they locate and extract useful data (e.g., tables, figures, maps, etc.) to construct the scientific databases.
However, it is nearly impossible for a single research team to manually collect and organize the scattered data from hundreds of thousands of documents, and the number is still increasing \cite{do-model-work}.
Although some larger research teams may have more workforce, without a well-designed computer-supported cooperative work (CSCW) platform, they still need to spend lots of effort to process a sufficient amount of literature and extract enough data for the scientific database.
Furthermore, a larger research team may face more team collaboration and coordination challenges such as synchronizing the work process and resolving conflicts.
Because of these challenges, constructing a scientific database using the data extracted from a large number of the literature often takes several years with a large workforce. This is a huge obstacle to the advancement of research.

With the development of artificial intelligence (AI), a few research teams have recently begun to explore building scientific databases with the help of AI.
Researchers have proposed several fully-automated information extraction systems (e.g., DeepDive \cite{zhang2015deepdive} and Fonduer \cite{10.1145/3183713.3183729}) with a goal to efficiently process documents in the PDF format and extract data.
Although these end-to-end fully-automated extraction systems may reduce the burden of manual document processing for researchers, they suffered the limitation of insufficient data extraction accuracy \cite{sun2022review}.
Furthermore, these systems require a huge amount of annotated data before training the AI extraction model, which is a well-established challenge and often infeasible in real-world to collect sufficient training data for multi-context deep-learning models \cite{goodfellow2016deep}.
Unlike many general annotation practices in computer vision and natural language processing \cite{chen2021goldilocks}, the data annotation and extract task in geoscience requires deep domain expertise \cite{karpatne2018machine}, making it impossible to obtain these labeled data using the general crowd-sourcing platforms \cite{snorkel}.
Consequently, researchers would have to pay extra effort for labeling and cleaning the training data to make the AI model work, which may take even more time than they manually extract data without using an AI.
We argue that instead of designing a fully-automated end-to-end solution, a human-AI collaborative and interactive system may be the solution to address these problems.
Geoscientists can perform the data extraction activity as they used to, and AI can train itself with these user-labeled data and then make suggestions to the user in the future.
Together, the human-AI team can accurately extract data at a much lower cost.

\begin{figure}[h]
  \centering
  \includegraphics[width=\linewidth]{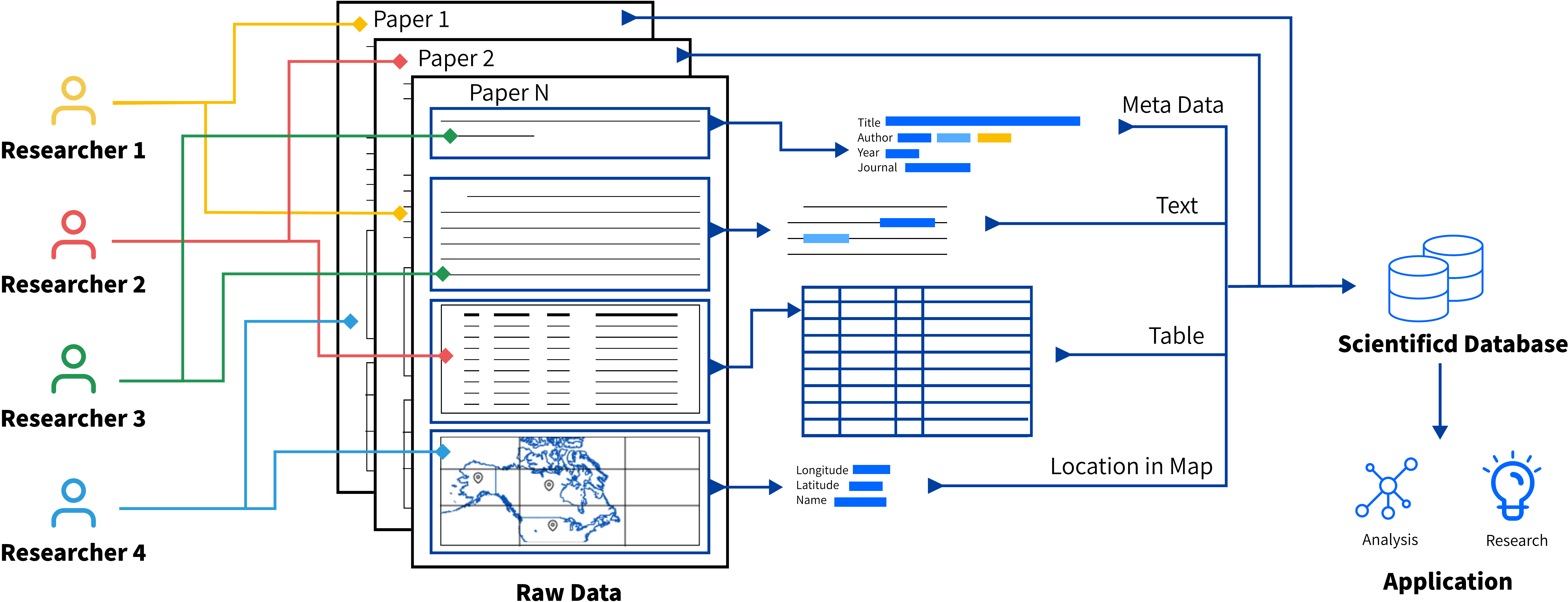}
  \caption{The collaborative data extraction from scientific literature for big data-driven geoscience research.}
  
  \label{fig:Workflow}
\end{figure}

To explore how geoscientists extract data from literature and then construct a scientific database, we recruited geoscientists from the Deep-Time Digital Earth (\DDE ) program \cite{oberhansli2020deep} and conducted a formative study to understand the problem space and identify user requirements.

The \DDE~program is a data-driven discovery program in geoscience with a goal to aggregate the geoscience data and to facilitate data-driven discovery for understanding Earth's evolution \cite{10.1093/nsr/nwab027}.
We designed and distributed a survey to \DDE~geoscientists, and then invited some of the respondents to participant our semi-structured interview sessions.
In total, we collect 119 questionnaires, of which 106 are valid,and interviewed 12 researchers with different roles in the team.
The formative study reveals the workflow of these geoscientists (Figure \ref{fig:Workflow}) and suggests three user needs:
\begin{itemize}
\item\textbf{N1:} Geoscientists need a more efficient way to collect and extract data from the documents.
\item\textbf{N2:} Geoscientists need a platform to help them conduct team collaborative data extraction.
\item\textbf{N3:} Geoscientists need a collaborative platform to organize their group research.
\end{itemize}

Based on these findings, we then designed \deepshovel, an AI-based collaborative data extraction platform to assist geoscience research teams to complete the work of data extraction, data aggregation, and scientific database construction.
\deepshovel~provides a user-friendly interface and experience design following the human-AI interaction design guidelines \cite{10.1145/3290605.3300233}, so that users even without any AI backgrounds can also work comfortably with it.
\deepshovel~has been deployed for one month and there are already 253 users from 36 geoscientist teams within the \DDE~program use it in a daily basis.
More than 240 projects and 46,000 documents have been processed for building scientific databases.
We further recruited 14 existing users from 9 teams and conducted a user study to understand their user experience of the system.

This paper makes the following contributions:
\begin{itemize}
    \item We conducted a needs-finding formative study to explore geoscientists' workflow and challenges in the process of data extraction from literature for building scientific databases, from which we proposed design suggestions for building the AI-based data extraction tools. 
    \item We developed an online team collaborative AI-assisted data extraction platform and have successfully applied it in the process of building scientific databases in geoscience research. 
    \item We proposed a human-AI team collaboration framework making the collaboration process between humans and the AI system more productive and having a better performance, which may generalize beyond the geoscientist data extraction use case.
    
\end{itemize}

\section{Related Work}

\subsection{Team Collaboration in Building A Scientific Database}

As big data-driven research becomes commonplace, an increasing amount of scientific databases are needed, and building a reusable database becomes a challenge for researchers.
In the field of geoscience, many researchers have pointed out that the process of data preparation is vital throughout the research process \cite{sun2022review}.
Many geoscientists think that building a system for data extraction access is necessary for a scientific database \cite{davydov2006detail}.

In the past, many researchers in geoscience tried to build an integrated platform from data collection and storage to data analysis, which promoted the sharing of geoscience databases and big data research.
Chronos \cite{cervato2005chronos} is a community facility addressing the needs of geoinformatics and providing simultaneous and seamless integration of hosted and federated databases with analytical and visualization tools.
Paleostrat \cite{snyder2008geoscinet} is designed as an infrastructure platform for Earth Science researchers and teachers, which serves the community by enhancing the research and education process.
However, they spent too much time designing schemas instead of trying to make a user-friendly interface for geoscientists to easily extract desired information, which makes it impossible for researchers to cooperate with their team in the data extraction process.
It leads to the conclusion that they do not have enough data to support their operation.

Some researchers noticed and analyzed the problems in team collaboration that existing platforms are facing.
\citet{hoeppe2021encoding} mentions that the large scientific databases often require large teams to extract and clean the data, and it is a complex task that needs to be done through CSCW.
\citet{schmidt2008taking} demonstrates that task distribution, allocation, and interrelating of ‘distributed individual activities' are some important issues in team collaboration.
\citet{steinhardt2014reconciling} think that these activities require the support of infrastructure.
\citet{finholt1997laboratories} concluded three core capabilities of technologies supporting scientific collaboration: linking people with people, linking people with information, and linking people with facilities.
However, they only provided some theoretical analysis and solution principles, and no implementation details and practicable platform have been proposed yet.

In this paper, we aim to reveal empirical understandings and build systems to support researchers' data extraction work with their teams.
Specifically, we investigated a research 1) to understand how researchers cooperated with their teams extracting data from scientific literature today,
2) to implement a system with AI assistance prototype to support their work, and 3) to explore researchers' feedback and design implications after they try it out.

\subsection{Data Extraction in Scientific Literature}

With the development of artificial intelligence, more and more researchers pay attention to the data and phenomena in the published literature.
Artificial intelligence methods are introduced in this process to help promote the collection of data from the published literature and build a scientific database in biology \cite{li2014biological}, chemistry \cite{swain2016chemdataextractor,rajan2021decimer}, and many other disciplines \cite{hong2021challenges,walker2022evaluation,decision-making-medical}.
In addition, there are also some related works focusing on literature retrieval and dissertation comprehension to extract information from scientific literature \cite{luan2018information,tkaczyk2015cermine,tkaczyk2014cermine}.

In the field of geoscience, GeoDeepDive \cite{GeoDeepDive} is a widely-used toolkit that adopts natural language processing (NLP) technology to process and analyze the literature end-to-end.
\citet{peters2017rise} use GeoDeepDive to study the evolution of stromatolites in shallow marine environments.
GeoDeepDive is a case study of DeepDive \cite{zhang2015deepdive} which stopped update as of 2017 and became a digital library named xDD \cite{xDD}.
Due to the use of supervised learning and end-to-end extraction methods \cite{govindaraju2013understanding}, the lack of labeled data has introduced the problem of insufficient data accuracy.
Moreover, these methods can only process text \cite{niu2012deepdive,AI-Labeling,AI-labeling-efficiency}, use NLP to analyze the tables with the content \cite{govindaraju2013understanding}, and do not store the tables as structured data for broader big data analysis.

However, most existing works on data extraction in the scientific literature using artificial intelligence require case customization and do not have a user-friendly user interface for researchers, leading to obstacles for researchers who do not have a artificial intelligence background \cite{lanius2021usability}.
\citet{wilkinson2016fair} demonstrate that scientific data needs to be findable, accessible, interoperable, and reusable.
Furthermore, data extracted by the customized case is not reusable, resulting in low data utilization.
We believe that retaining structured data can increase data reusability and avoid duplication of data extraction, which is why we consider the idea of Human-AI Collaboration.
We study the current workflow in which geoscientists manually extract data from the literature.
Based on this workflow, \deepshovel~adds the use of AI assistance and human-AI collaboration thinking into the data extraction process.
\deepshovel~can improve efficiency and data reusability and avoid the lack of precision caused by automatic methods.

\subsection{Scientific Document Processing and Human-AI Collaboration}
For scientific literature parsing, there are some toolkits and packages like Grobid \cite{GROBID}, Science Parse \cite{tkaczyk2018machine} and PDFFigures 2.0 \cite{clark2016pdffigures}.
However, on the one hand, they can only decompose the paper, but still cannot form data that can be used for research.
On the other hand, the lack of graphical interfaces means it is difficult for researchers who are only average computer users to use these tools and carry out further data processing easily.
Some other interactive document processing tools take ease of use and user-friendliness into account.
ABBYY FineReader PDF \cite{finereaderpdf} is a desktop application for processing PDF documents.
It provides an OCR tool that makes the content of a PDF document editable.
The extraction processes of text, pictures, and tables in the PDF document is mixed so that the user needs to deal with all the information in process, which causes information redundancy and makes it difficult for users to focus on the data they need.
TableLab \cite{wang2021tablelab}, which is an interactive table extraction system, can help people to extract and train a customized model.
However, due to the focus model training, TableLab's user interaction process presents a mixture of table structure and content recognition.
It means that users need to pay attention to the correctness of both table structure and content recognition at the same time.
Such information overload will increase the difficulty of users' decision-making and introduce the problem of model overload \cite{decision-making,mixed-initiative-user-interfaces}.
Moreover, these tools are not designed for scientific literature.

\deepshovel~provides users with a friendly graphical user interface and a well-design interactive extraction process between humans and AI to prevent information overload and model overload.
Users can extract the structured data they need, and the consideration of human-controlled decision making ensures the accuracy of the data.
Such a human-AI collaboration framework and functional design with team collaboration can help researchers without relevant technical backgrounds quickly use \deepshovel~to extract data from the literature and quickly invest in research with their teams.

\section{User Research and Requirement Analysis}
In our project, we follow the design study methodology \cite{olson2014ways}. 
As it suggests, we start analyzing the real-world problems of domain experts and working on creating a system to solve these problems. 
This section describes how we conduct the user research and what we learn from the user research. 
In the task definition stage, to understand the relevant technical background of the domain experts and the data extraction involved in their research questions, we used questionnaires to investigate 119 related potential users and selected 9 groups of users among them for 60 minutes of in-depth interviews.
Then we analyzed and summarized the commonalities of the user groups and their main tasks and difficulties.

\subsection{Questionnaire Survey}\label{Q}

We choose to use questionnaire surveys \cite{muller2014survey} to study, which brings better user characteristics understanding, user requirements acquisition, and system design.
Specifically, we hope to use the questionnaire to understand the current status of work in the field of earth sciences on the task of constructing subject-specific databases.
We divided the questionnaire into three main parts:
\begin{itemize}
    \item Basic Information
    \item Task- related Information
    \item User's understanding of computer technology
\end{itemize}
Based on the above disassembly of the question, we designed the initial version of the questionnaire and invited potential interviewees to conduct face-to-face interviews to clarify and eliminate the cognitive bias that the interviewee may have.
Finally, we determined the topic and the questions of the questionnaire as shown in Table \ref{tab:Questionnaire}.

Since our research is aimed at a specific field, we did not choose the probability-based or random sampling method for survey research. 
We used non-probability-based methods, invited relevant researchers in the \DDE~program, took a voluntary form to participate in our questionnaire survey, and combined in-depth interviews to conduct in-depth research.
We used an online questionnaire to overcome geographic barriers and covered more researchers in different regions.
A total of 119 questionnaires were obtained, of which 106 were valid.

\begin{table}
    \centering
    \begin{tabular}{ll}
    \toprule
     No. & Questions\\
    \midrule
    \multicolumn{2}{l}{\textbf{Part 1: Basic Information}}\\
    Q1 & Gender\\
    Q2 & Career Position\\
    Q3 & Research Field in Geoscience\\
    \midrule 
    \multicolumn{2}{l}{\textbf{Part 2: Task- related Information}}\\
    Q4 & The progress of Data extraction\\
    Q5 & Research Team size\\
    Q6 & Team composition\\
    Q7 & Tools using in data extraction\\ 
    Q8 & Data source used\\
    Q9 & Data source format and proportion\\
    Q10 & Method to process PDF files\\
    Q11 & Distribution and proportion of data in the file\\
    Q12 & Number of database fields\\
    Q13 & The number of documents needed to build the database\\
    Q14 & Personal participation in data collection\\
    Q15 & Current Workflow of data extraction\\
    Q16 & Time of Single PDF processing\\
    Q17 & Estimate the time required for the entire collection\\
    \midrule 
    \multicolumn{2}{l}{\textbf{Part 3: Understanding of computer technology}}\\
    Q18 & Understanding of programming\\
    Q19 & Kinds of tasks be accomplished by programming\\
    Q20 & Understanding of artificial intelligence\\
    Q21 & Data set preparation of artificial intelligence task\\
    Q22 & Understanding of data labeling\\
    Q23 & Kinds of tasks served by data labeling\\
  \bottomrule
    \end{tabular}
    \caption{Questions in the questionnaire.}
    \label{tab:Questionnaire}
\end{table}

\subsection{In-depth Interview}\label{I}

Based on the questionnaire survey, in order to know details of geoscientists' research and team cooperation, we conducted the in-depth interviews as semi-structured interviews \cite{longhurst2003semi}. 
We organized the semi-structured interview with a framework of questions about their research interests, teamwork, and usage of the data they collected to explore their needs during the workflows.
We invited 12 users for 30-60 minutes interviews from the questionnaire survey participants who have different roles in their research team.
The interviewees' research projects and roles of their research team are shown in Table \ref{tab:Interviewees}.
\begin{table}[]
    \centering
    \resizebox{1. \linewidth}{!}{
    \begin{tabular}{cccll}
    \toprule
    \makecell[c]{Group\\No.} & \makecell[c]{Participant\\No.} & Gender & Research Projects & Roles\\
    \midrule
      G1   &  P01 & Male & Magmatic Migration & PhD Student\\
      G2   &  P02 & Male & Geomagnetism and Geoelectromagnetism & Associate Professor\\
      G2   &  P03 & Male & Geomagnetism and Geoelectromagnetism & PhD Student\\
      G3   &  P04 & Male & Paleoclimatology & Associate Professor\\
      G4   &  P05 & Male & Geochronology and Structural Geology & Professor\\
      G5   &  P06 & Female & Paleontology & Associate Researcher\\
      G6   &  P07 & Male & Structural Geology & PhD Student\\
      G7   &  P08 & Female & Evolutionary Biology and Dinosauria & PhD Student\\
      G7   &  P09 & Male & Evolutionary Biology and Dinosauria & PhD Student\\
      G8   &  P10 & Male & Carbonate Sedimentology & Postdoctoral Researcher\\
      G9   &  P11 & Female & Global Detrital Zircon Database & Full-time Data Entry Clerk\\
      G9   &  P12 & Male & Global Detrital Zircon Database & Full-time Data Entry Clerk\\
  \bottomrule
    \end{tabular}
    }
    \caption{Demographics of interviewees.}
    \label{tab:Interviewees}
\end{table}
\subsection{Results}

\subsubsection{Challenges for Data Extraction}
According to the results of the questionnaire, the main challenge is to help geoscientists without backgrounds in computer extract structured data from unstructured data sources, especially PDF documents.

The PDF is the most significant proportion of the document formats that geoscientists pay attention to.
Due to the difference in encoding, version, and source of PDF documents, the structure of internally stored digital information is chaotic and can not be processed automatically by machines.
Especially for some PDF files scanned from paper, the scanning quality dramatically affects the results of automated processing.
All the interviewees consider such errors unacceptable, as they will directly impact the research results.

In addition, the results of the questionnaires and interviews show that the most valuable data are located in the tables in the article.
However, tables in PDF are usually not reproduced very well by ordinary PDF editors, which makes data collection very difficult.

Another problem introduced by completely manual copy-pasting of data is the proofreading of the data.
The interviewees P01, P11, and P12 say that they need to spend much time proofreading the data and matching them with the meta information of the corresponding source article in the database.

\subsubsection{"We don't know much about programming for data extraction, and we can't utilize state-of-the-art AI models without graphical user interfaces"}\label{excel}
\begin{figure}
    \centering
    \includegraphics[width=\linewidth]{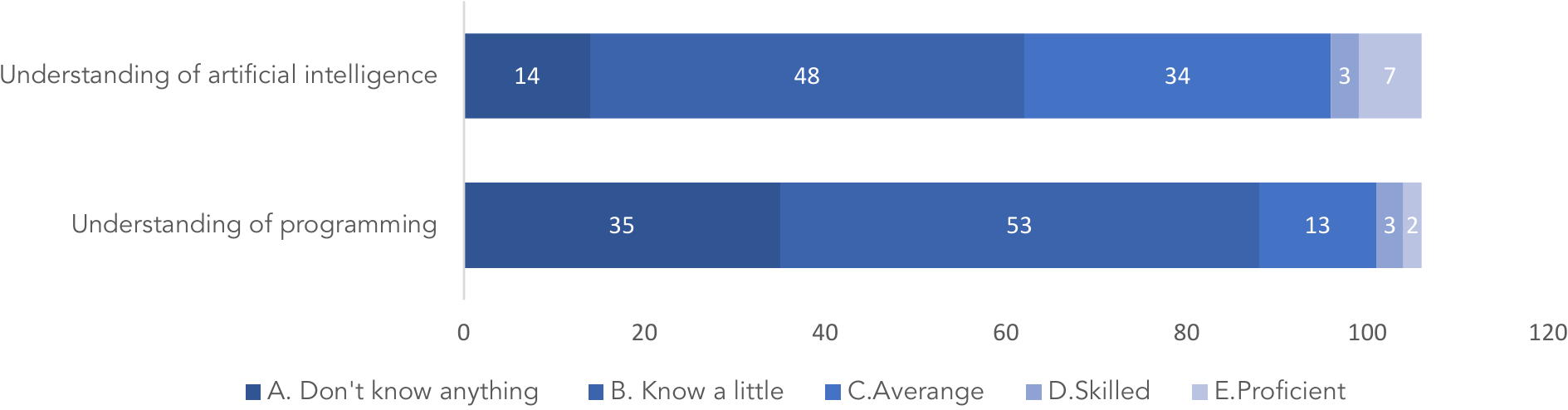}
    \caption{Results of Q18-Understanding of programming \& Q20-Understanding of artificial intelligence.}
    \label{fig:understanding_of_ML}
\end{figure}

\begin{figure}[!t]
\centering
\subfloat[Q7-Tools using in data extraction.]{\includegraphics[width=1.\linewidth]{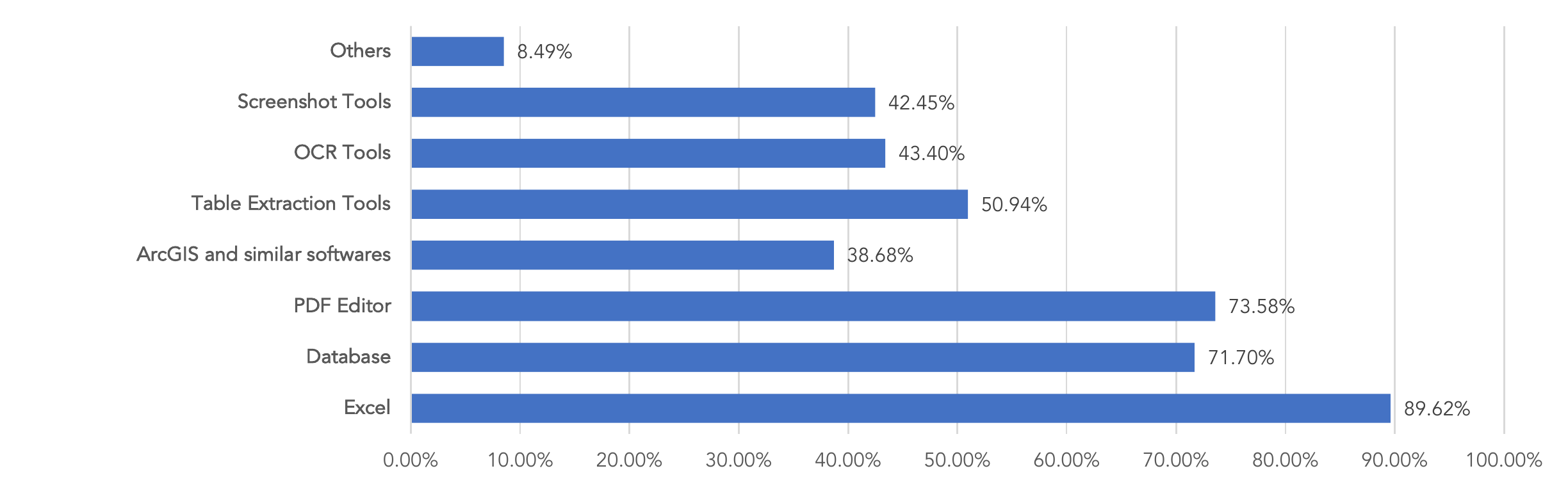}
\label{fig:temporal-observation-a}}
\hfil
\subfloat[Q10-Method to process PDF files.]{\includegraphics[width=1.\linewidth]{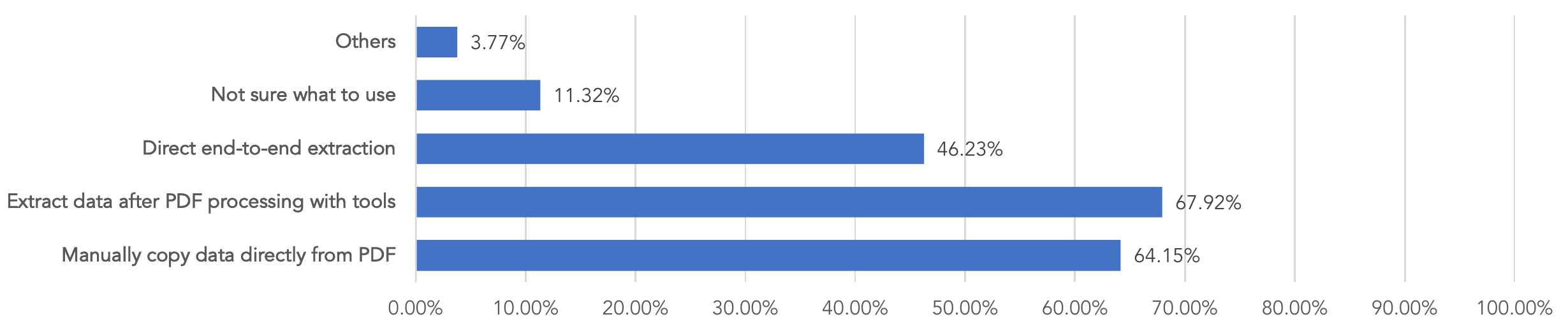}
\label{fig:temporal-observation-b}}
\caption{Results of Q7 \& Q10.}
\label{fig:tools_result_questionnaire}
\end{figure}

From the survey research and in-depth interview, we conclude the typical profile of our users that they are average computer users with no / low programming skills.
Only P07 is a proficient programmer and can use Python scripts to process the data, but he still mentioned that "Being unfamiliar with artificial intelligence makes me unable to deal with the data in pictures and charts in the papers efficiently".
We can find that most participants of the survey also lack understanding of programming/artificial intelligence (Figure \ref{fig:understanding_of_ML}).
Furthermore, some interviewees also report that they always encounter difficulties in data extraction and processing due to the lack of programming skills.
We can learn from the questionnaire results that our users have to use a combination of a series of tools to accomplish a task (as shown in Figure \ref{fig:tools_result_questionnaire}).
They usually use some OCR tools with graphical user interfaces to process PDF documents to make them editable and then manually copy-paste the data they find into Excel.
We found that an online collaborative application with a user-friendly graphical user interface is critical for geoscience research teams.

\subsubsection{Difficulties of Team Collaboration}
\label{Team_collaboration}
Interviewees mentioned that they worked as a team to accomplish data extraction tasks and indicated some problems in team collaboration remain unsolved.

The first difficulty is the storage of data sources related to the distribution of team tasks and the advancement of tasks.
We notice in the questionnaire results that respondents generally believe that more than 1,000 articles are needed to construct a scientific dataset.
Moreover, we learn from the interviewees that the current methods of researchers sharing literature within teams are still relatively primitive (e.g., copying files and excel sheet records).
Using these methods is quite laborious when sharing a large number of files and can cause huge risks of data errors and version conflicts.
We believe that building a scientific database is a close-cooperation work but does not have a CSCW system to support it.

Another finding is that Excel is an essential tool throughout the process.
Once the data is extracted from the PDF, it will be stored in Excel as we mentioned in \S\ref{excel}.
P01 and P13 mention that each team member stores data from his documents into a local Excel file and then merges it with other people's Excel data files in the team.
There are great similarities in the ways of collaboration among different teams, which are all primitive and loose.

\subsubsection{Summary of the Workflow}\label{Task}
\begin{figure}
    \centering
    \includegraphics[width=\linewidth]{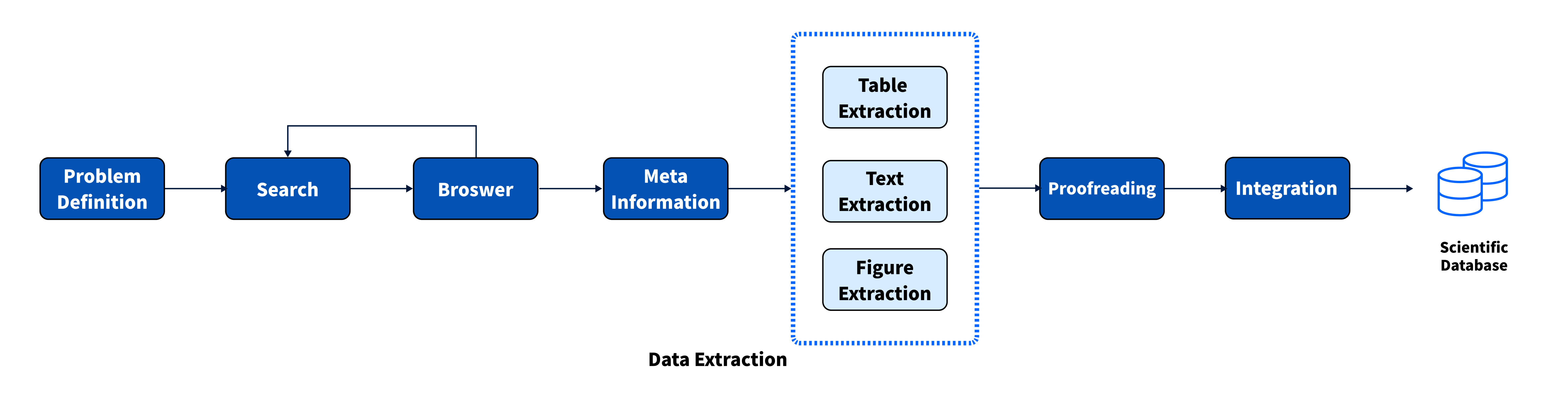}
    \caption{The team's workflow of collecting and building a database from geoscience literature.}
    \label{fig:data_extraction}
\end{figure}

Although researchers from different sub-fields of geoscience may study different research questions and focus on different data, their research processes and workflows are basically the same.
In this paper, we focus on the data extraction process of their research, as shown in Figure \ref{fig:data_extraction}.

Based on the results of interviews and questionnaires, we described the data extraction process and defined the tasks in the process.
These tasks are grouped and detailed by the main workflow steps in the following list:
\begin{itemize}
    \item \textbf{T1-Problem Definition:} Define the research problem and the structure of the scientific database that needs to be built,
    \item \textbf{T2-Search:} Search for the paper that may contain the data about the research problem,
    \item \textbf{T3-Browser:} Quickly browse the article to find data that is needed,
    \item \textbf{T4-Meta Information Extraction:} Record the literature's meta information for tracking the data,
    \item \textbf{T5-Detail Data Extraction:} Extract data from different parts of the literature,
    \SubItem{\textbf{Data extraction from table:} Get the data in the table and fill in the Excel file prepared in advance cell by cell,}
    \SubItem{\textbf{Information extraction from text:} Search the full text with keywords to locate the data, fill in the data in the Excel file after finding it, and repeat until all the data is found,}
    \SubItem{\textbf{Information extraction from figure:} Restore the corresponding information from the charts, such as obtaining the latitude and longitude of a marked point from the map,}
    \item \textbf{T6-Proofreading:} Check and proofread the data to ensure the data is accurate (usually done by the member and the team manager together),
    \item \textbf{T7-Data Integration:} Integrate the data extracted from each paper (usually stored in a bunch of Excel files) into the final dataset.
\end{itemize}

We also summarized the usual structure of teams building scientific databases and the related task for each role as shown in Table \ref{tab:roles}.
We also want to mention that some interviewees said they built the scientific database individually because not much literature needed to be processed.
Meanwhile, in some small teams, the responsibility of the role of "Member" might be taken by the role of "Manager" due to the lack of manpower.

\begin{table}[]
    \centering
     \resizebox{0.8 \linewidth}{!}{
    \begin{tabular}{c|l|c|c}
    \toprule
       Role  &  Responsibility &  Possible person(s) for this position & Participate in the task flow  \\
       \midrule
       \makecell[c]{Owner} & \makecell[l]{Determine the research topic, \\Collect relevant literature,\\ Determine database structure,\\ Check the quality of the data.} & Research team leader & T1 \& T2\\
        \midrule
        Manager & \makecell[l]{Distribute tasks,\\ Manage task progress,\\ Proofread the extracted data, \\Integrate data. }& \makecell{Senior researchers or\\ Senior members of the team} & T3, T6 \& T7\\
        \midrule
        Member  &  Specific data extraction.   & \makecell[c]{Junior researcher or\\ Professional trained data entry clerk} & T4, T5 \& T6   \\
        \bottomrule
    \end{tabular}
     }
    \caption{Teams structure and related tasks for each role.}
    \label{tab:roles}
\end{table}
\subsubsection{Design Requirements}
In this workflow, we found that there are three levels of requirements influencing the efficiency and experience:
\begin{itemize}
    \item \textbf{R1: } Quickly and accurately extract data from PDF files and form structured data,
    \item \textbf{R2: } Help proofread and integrate the data extracted by each person to build a database in multi-person collaborative extraction,
    \item \textbf{R3: } Share tasks' resources (raw data) and tasks' progress across the team.
\end{itemize}

\begin{figure}[h]
  \centering
  \includegraphics[width=0.8\linewidth]{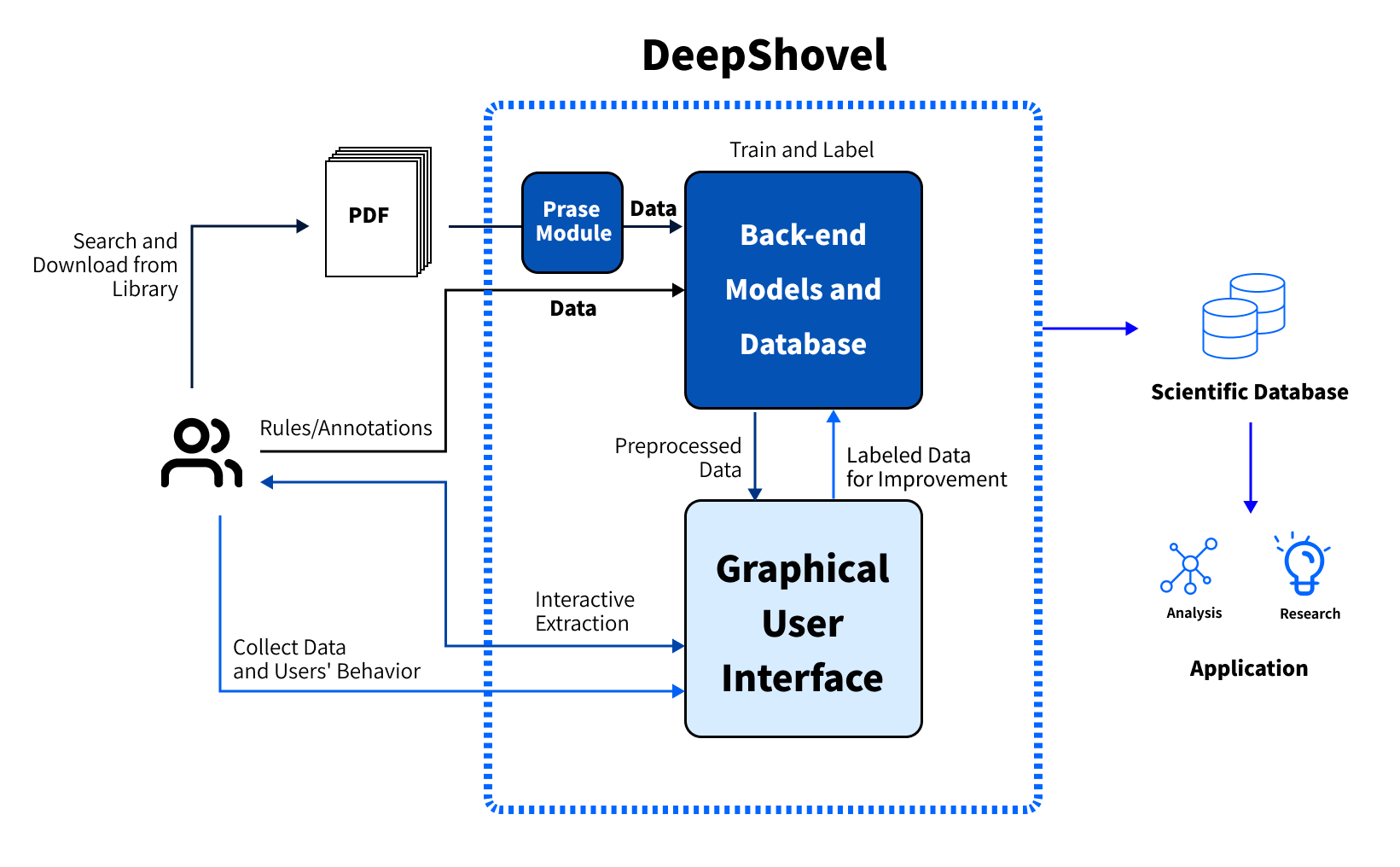}
  \caption{System overview.}
  \Description{}
  \label{fig:system_overview}
\end{figure}

\section{System Design}
Based on user research and requirement analysis, we solve these problems using a human-AI collaboration system design to extract data easily from PDF and have a better team cooperative experience.
For each task mentioned in \S\ref{Task}, we design and implement some artificial intelligence modules. 
According to human-AI collaboration thinking, the system can collect the data for artificial intelligence model training while assisting humans in finishing the tasks. 
Such collaboration motivates people to participate and allows the machine to obtain enough information to improve. The models implementation details are shown in \S\ref{data_extraction_model}.

\subsection{System Overview}
As shown in Figure \ref{fig:system_overview}, \deepshovel~consists of: 
(1) a interactive graphical user interface (see Figure \ref{fig:UI_overview}) including data extraction, document management, team management and data integration (D in Figure \ref{fig:UI_overview}); 
(2) a back-end parse module to pre-process the PDF format files;
and (3) some back-end artificial intelligence models supporting data extraction and integration functions.

\subsection{Human-AI Collaboration for Data Extraction}
\label{Human-AI_Collaboration_for Data_Extraction}

When users open a file from Project File List to start their work, they will enter the data extraction interface (Figure \ref{fig:meta}). 
In the data extraction interface, users can switch the different tabs (e.g., Meta, Text, Table, and Map) in the area F1. 
The details of each function are in the following sections.

\subsubsection{Meta Information Extraction}
\begin{figure}
    \centering
    \includegraphics[width=0.9\linewidth]{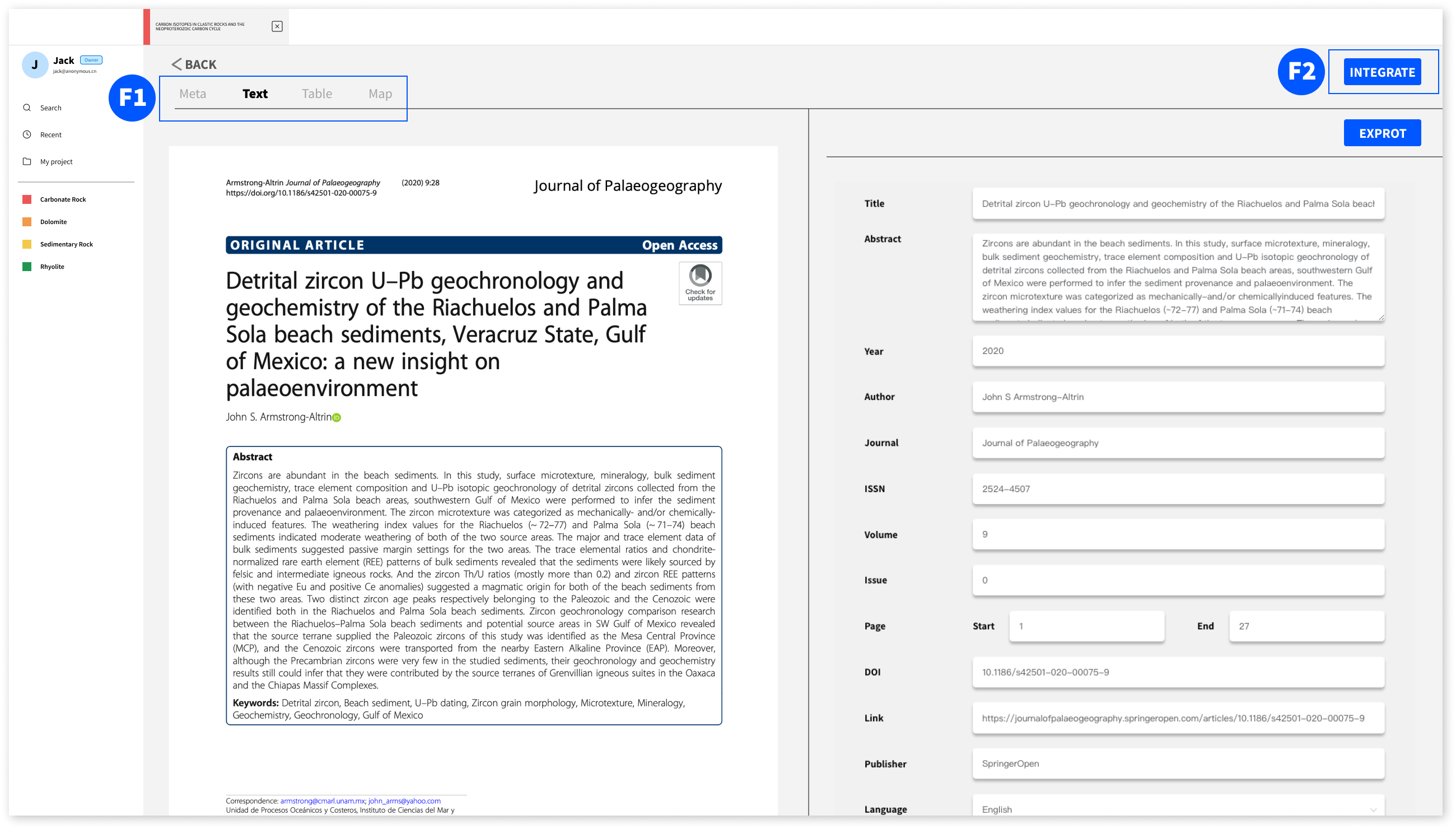}
    \caption{UI of meta information extraction.}
    \label{fig:meta}
\end{figure}

For the task \textbf{Meta Information Extraction}, we develop a module to automatically extract the title, author(s), journal/conference, and other meta information from the PDF file. 
As shown in Figure \ref{fig:meta}, users can edit and save the meta information that can be joined to the output dataset (refer to \S\ref{fusion}).

\subsubsection{Text Extraction and Annotation}
\begin{figure}
    \centering
    \includegraphics[width=0.9\linewidth]{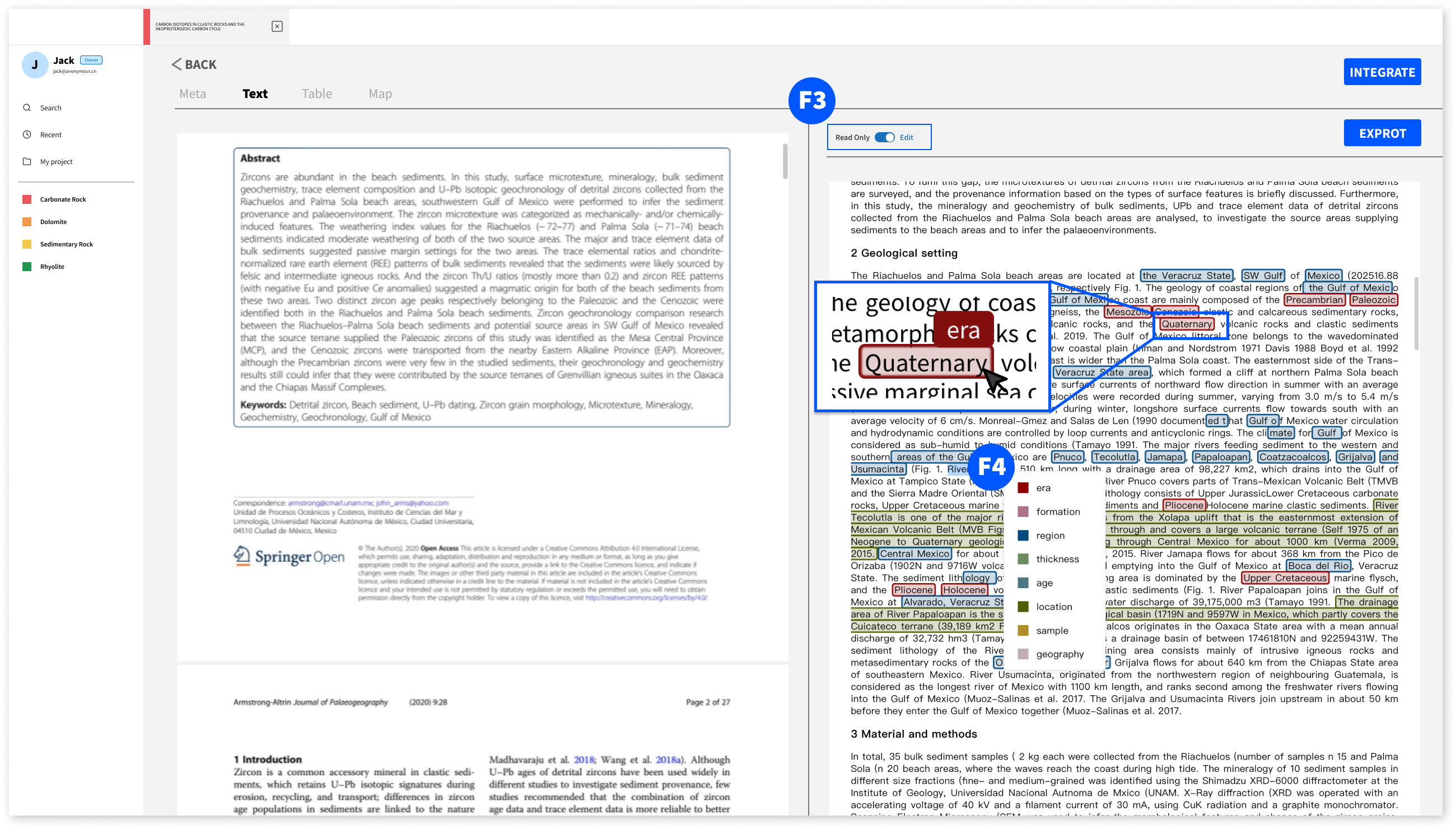}
    \caption{UI of text extraction.}
    \label{fig:text}
\end{figure}

We use weak-supervision learning models and rules to help highlight the focused keywords \textbf{in texts} and the samples' features to help them add these words into the database.
For example, we have a dictionary of eras' names to highlight the era mentioned in the PDF.
Users can also annotate a keyword via mouse selection when they switch to the edit mode (F3) and select a label (F4) as shown in Figure \ref{fig:text}. 
The keywords can be added to the output database (refer to \S\ref{fusion}).
Users can choose to show or hide some labels, which are set at the project level as shown in Figure \ref{fig:text_setting}.

\subsubsection{Table Extraction}
\begin{figure}
    \centering
    \includegraphics[width=0.9\linewidth]{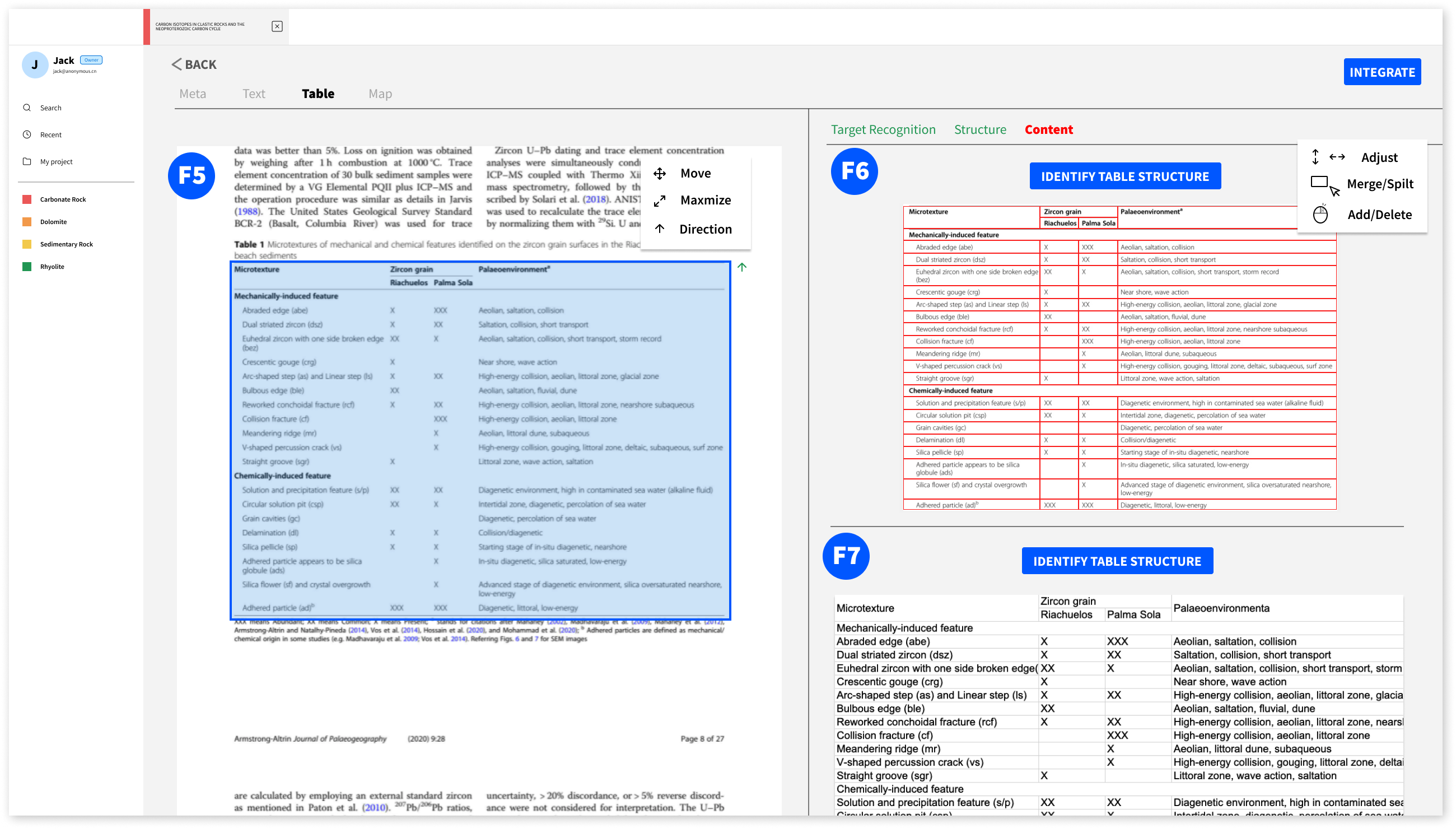}
    \caption{UI of table extraction.}
    \label{fig:table}
\end{figure}

To help user extract the data in the table, we develop the \textbf{Table Extraction} function (Figure \ref{fig:table}).  
In this part, we separate the task into three steps: 
1) Locate the Table with the assistance of AI;
2) AI recognizes the table's structure, and users assist for a better result;
3) AI recognizes the table's content, and users can edit for final accuracy.
In each step, the artificial intelligence models we design will help people to easily get the result and collect the users' adjustments for model training(see Figure \ref{fig:table_extraction_interaction}). 
From the user's perceptive, the first step is adjusting where the table is in the F5 area or drawing a new area as a table, then starting the recognition of structure. 
The next step is to adjust the structure that the system advised (F6).
The system provides `add and delete column/row' and `merge or split cell function'.
After structure recognition, users can start the content recognition and edit the content in each cell (F7).

\subsubsection{Map Recognition and Location Extraction}
\begin{figure}
    \centering
    \includegraphics[width=0.9\linewidth]{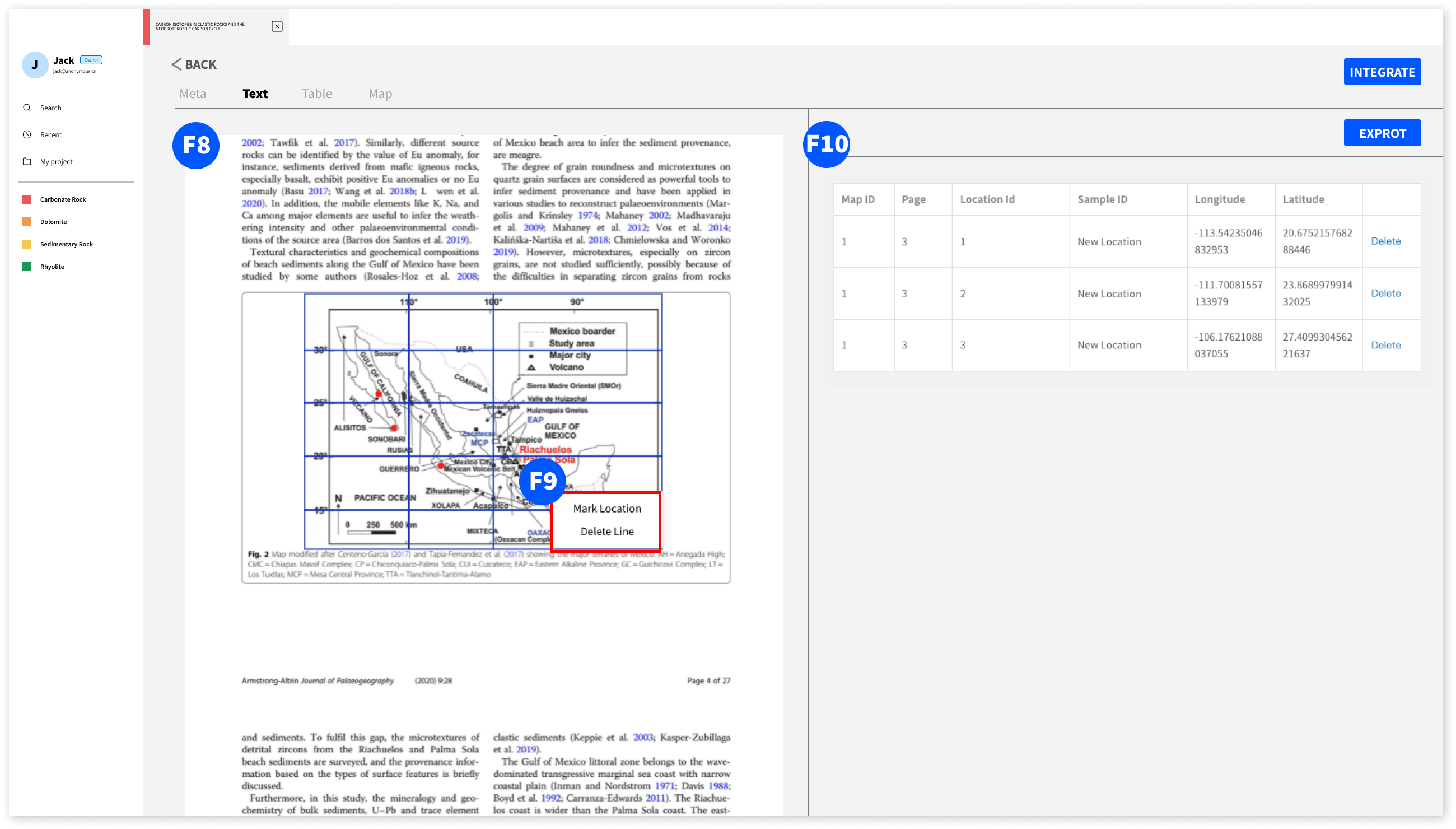}
    \caption{UI of map recognition and location extraction.}
    \label{fig:map}
\end{figure}

For collecting the location of a sample, we provide a module that can \textbf{recognize maps} and calculate the latitude and longitude of each point on the map (Figure \ref{fig:map}). 
Users can draw an area (F8) that contains the map and mark a point by right click (F9). The latitude and longitude will automatically be saved in the table (F10) as shown in Figure \ref{fig:map} and can be joined to the output dataset (refer to \S\ref{fusion}).

\subsection{AI-Assisted Team Collaboration}\label{fusion}
\begin{figure}
    \centering
    \includegraphics[width=0.9\linewidth]{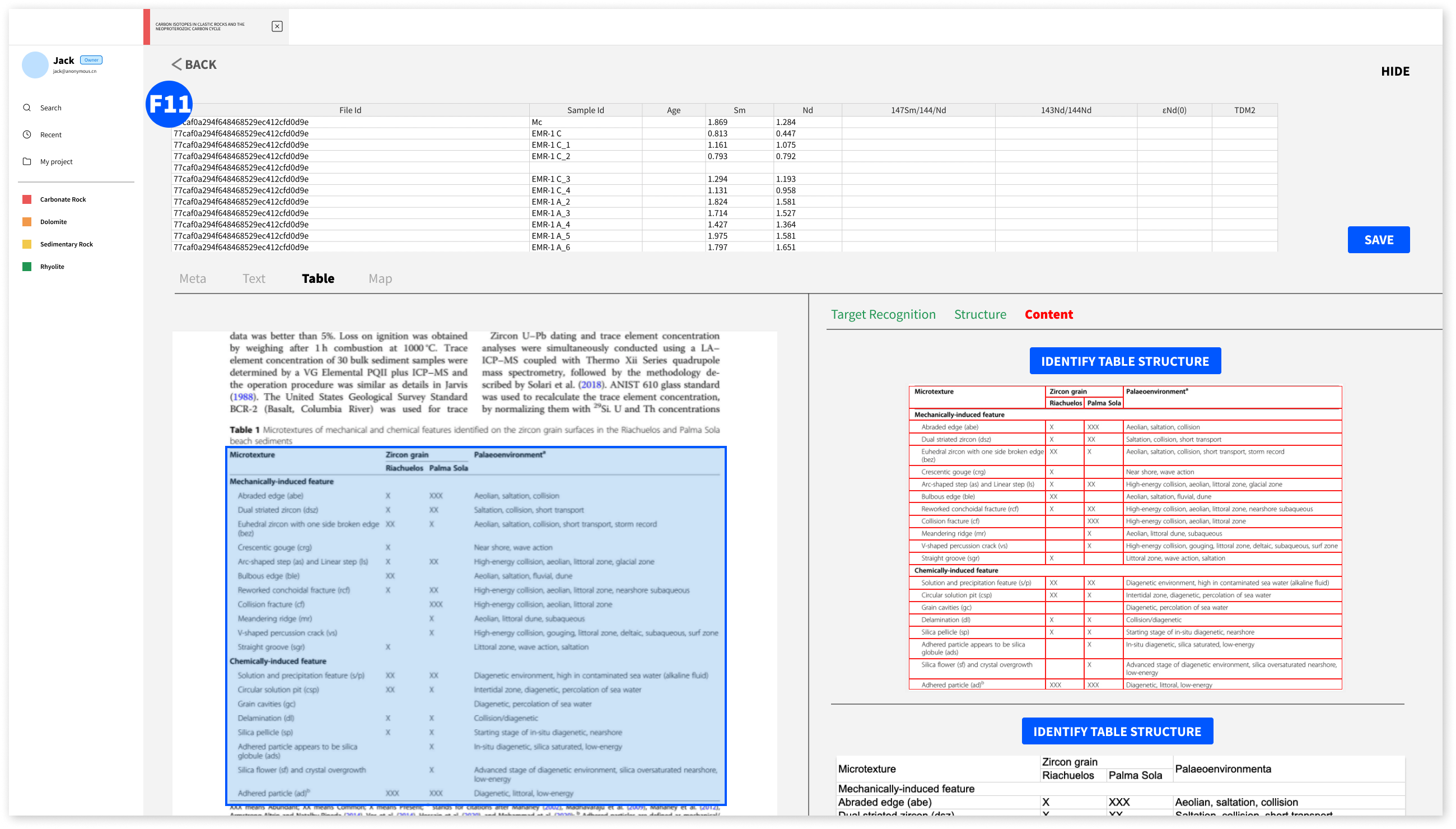}
    \caption{UI of data integration.}
    \label{fig:integrate}
\end{figure}

\begin{figure}[!t]
\centering
\subfloat[The settings of text extraction.]{\includegraphics[width=0.44\linewidth]{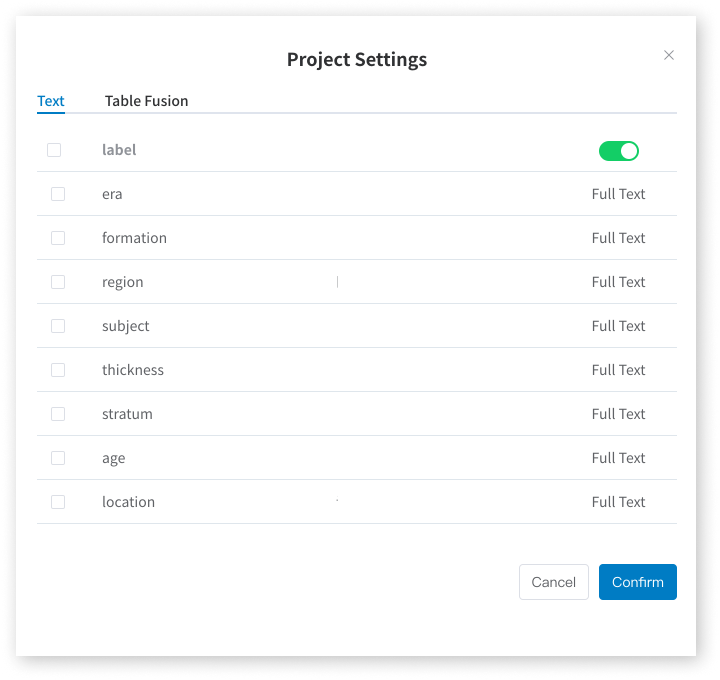}
\label{fig:table_fusion_setting}}
\hfil
\subfloat[The settings of data integration.]{\includegraphics[width=0.44\linewidth]{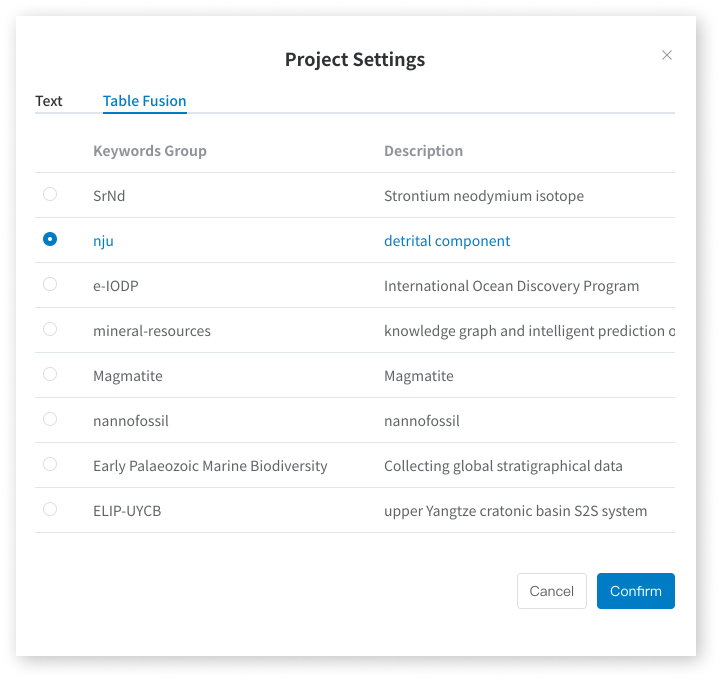}
\label{fig:text_setting}}
\caption{The project settings.}
\label{fig:settings}
\end{figure}

After the data is extracted step by step, it needs to be integrated into a table to establish a database finally. 
It involves how the data extracted by everyone in the team can be put into a summary table faster. 
We designed a single file integration and project-level integration with the assistance of AI to adapt to different teamwork modes.
The user needs to set the header of the total table on the project page (Figure \ref{fig:table_fusion_setting}).
Then in the data extraction interface (Figure \ref{fig:meta}), when users click the Integrate button (F2), the back-end model will process the data in each part, including meta information, tables, location in maps and texts. 
The result are shown at the F11 area in Figure \ref{fig:integrate}.

After all the data in a single file has been integrated into a file-level summary table, the user can integrate the summary table of each file into a project-level summary table at the Files List interface (F12 in Figure \ref{fig:file_list}), and the result will be automatically downloaded.

\subsection{User-Document Management Design for Team Collaboration}
As we mentioned in \S\ref{Team_collaboration}, users need to process a large number of files and need to distribute files to team members for data extraction.
Therefore, we provide a user-document management with graphical user interface that can be easily used.

\subsubsection{Project and File Management}
\begin{figure}
    \centering
    \includegraphics[width=0.9\linewidth]{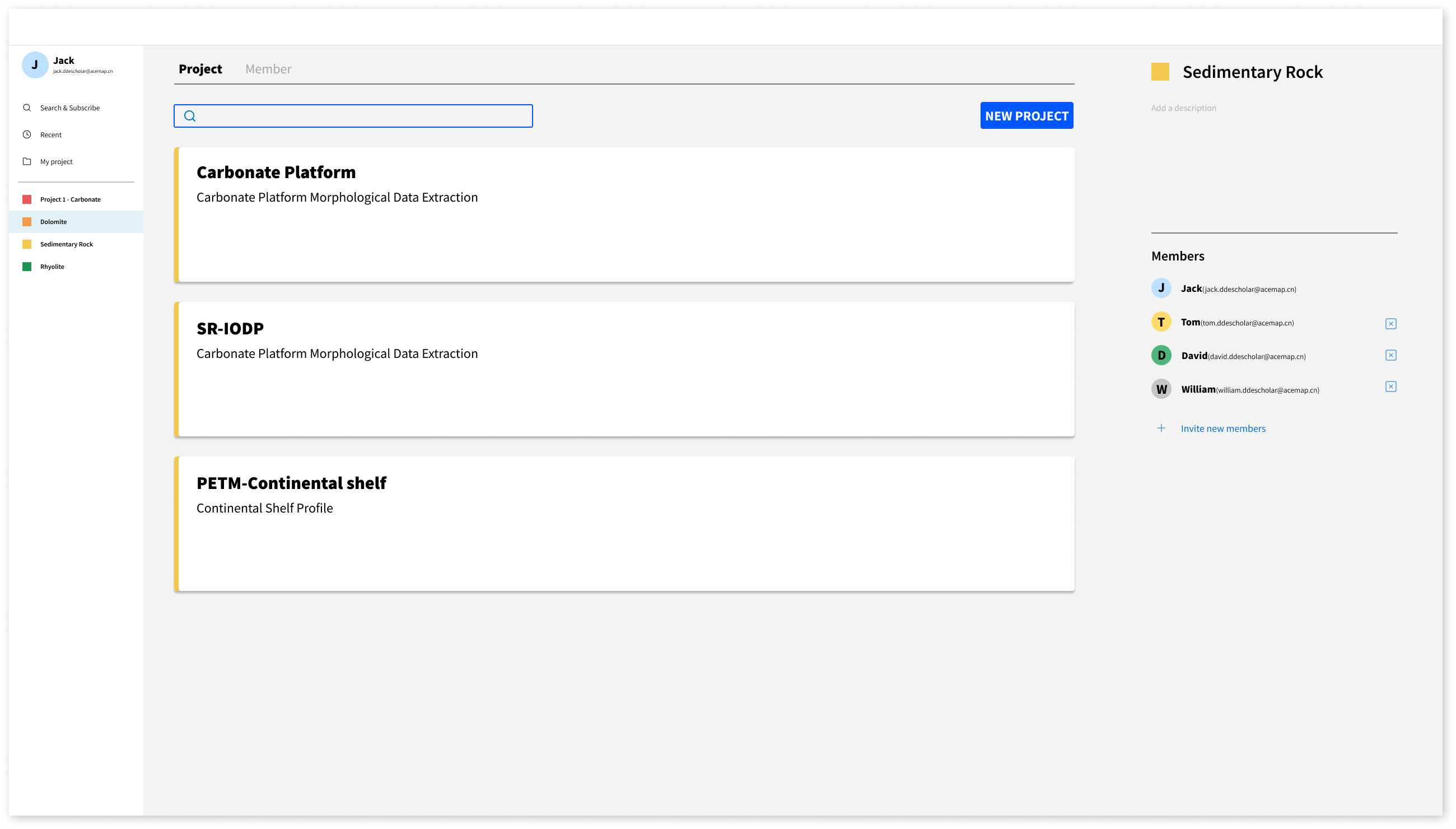}
    \caption{UI of project list.}
    \label{fig:project_list}
\end{figure}

\begin{figure}
    \centering
    \includegraphics[width=0.9\linewidth]{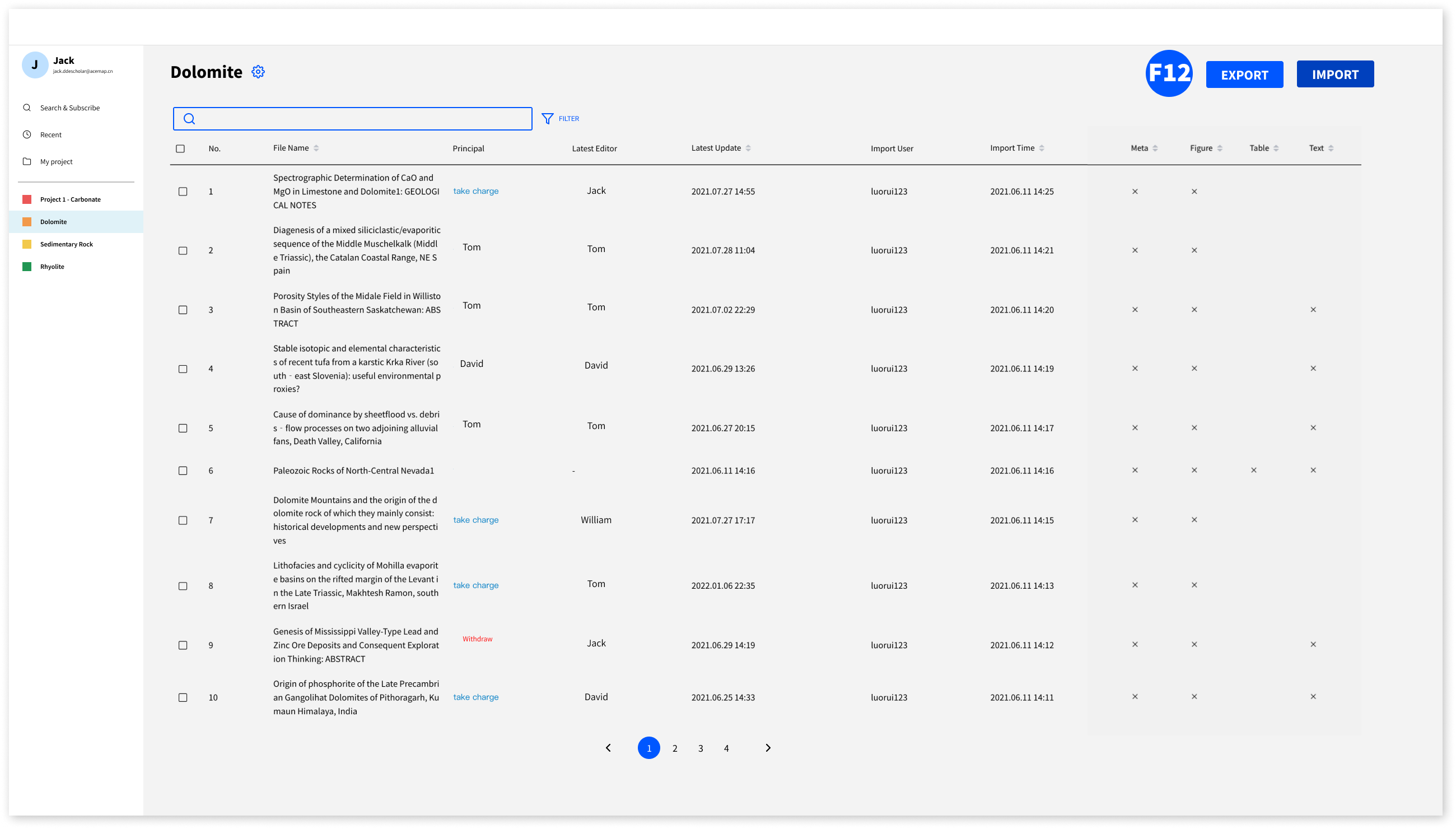}
    \caption{UI of files list.}
    \label{fig:file_list}
\end{figure}

In order to realize the management of projects, we designed the projects list interface to display the relevant information of each project, as shown in Figure \ref{fig:project_list}.
Each project has a file list (see Figure \ref{fig:file_list}), which will show who uploaded the file, who was the last editor, the upload time and last edit time, and whether the file has a principal (refer to \S\ref{principal}).
Users can change the project settings, including the text labels, the export dataset headers, and the project description.
Considering that the dataset may contain several headers, we provide batch edit for convenience.

\subsubsection{Roles Settings and Cross-team Cooperation}
To help researchers manage their team in data extraction work, we preset three team roles in the system design: Owner, Manager, Member.
The user permissions of the three roles are shown in Table \ref{tab:roles}.

\begin{table}[]
    \centering
     \resizebox{1. \linewidth}{!}{
    \begin{tabular}{c|c|c|c|c|c}
    \toprule
       Role  &  Add/Remove Manager & Add/Remove Member &  Add/Delete Project & Import File & Project Settings  \\
       \midrule
        Owner   & $\surd$ & $\surd$ & $\surd$ & $\surd$ & $\surd$\\
        Manager & -       & $\surd$ & $\surd$ & $\surd$ & $\surd$\\
        Member  & -       & -       & -       & $\surd$ & - \\
        \bottomrule
    \end{tabular}
     }
    \caption{The user permissions of each role.}
    \label{tab:roles}
\end{table}

According to the users' description of their current team structure and cooperation in the formative study, the users intend to ensure the original data's controllability and distinguish different research projects (the same team may carry out multiple projects).
Therefore, in team management, we ban the modification of the project by the Member role to prevent the original data from being modified.
At the same time, to ensure the rigor of the output dataset, we only open the modification of project settings to Manager and Owner.

Our cross-team collaboration design is based on research in the CSCW community on patterns of information exchange among teams in research work \cite{velden2013explaining}, and on respondents' descriptions of this behavior.
Information exchange is required among teams due to knowledge sharing.
In order to meet the common scenario of cross-team collaboration in research, we allow users to join different teams. 
Team member removal does not affect his/her past actions.

\subsubsection{Lock and Principal Mechanism}\label{principal}
Online collaboration systems often encounter the situation of simultaneous editing. 
Considering the particularity of system functions and the interaction process with the back-end model, we designed a file locking mechanism to prevent more than one user from operating on the same file.
Based on file lock, we implement a principal mechanism. 
Users can click the "Take Charge" button in the list to choose to be the "principal" in charge of a file.
Then the file can only be operated by the principal user, and other users can only read it. 
The user can release the file permission at any time.

\subsection{System Implementation Details}
\subsubsection{Web Application}
The front-end interactive single-page web application of \deepshovel~is developed in Vue.js and hosted with Nginx.
The web-based design of \deepshovel~gives it the ability to run in the web browsers on a variety of platforms including desktops, laptops, tablets, and smartphones.
The use of Vue.js and the design of single-page-application bring extreme load speed similar to native apps and consistent user experience across devices and platforms.

The back-end API service of \deepshovel~is implemented with Python and FastAPI framework.
The asynchronous coding design makes it possible to achieve higher concurrency with a minimal resource occupation so that it can support more users at the same time.
We adopt a master-slave backup MySQL database to store documents and extracted data, which ensures data security and efficient reading and writing.
In terms of user system security, we only store and bcrypt hashed passwords to ensure that users' plaintext passwords will not be stored and leaked.
Moreover, the HTTPS protocol is applied to the whole system of \deepshovel~to ensure the security in network communication.

\subsubsection{Document Management \& Retrieval}
All the literature uploaded into \deepshovel~are all automatically parsed with Grobid \cite{GROBID} and Science Parse \cite{tkaczyk2018machine}.
The meta information of papers (e.g., Title, Author List, Abstract, Venue, and Year) is extracted and indexed with Elasticsearch.
Then all the fields could be utilized for searching and retrieving the documents.
Moreover, to better browse and manage literature, the document list could be filtered with the principal user as well as the import user and sorted by title, import time, and latest update time.
To go further, each user could get ``My File List'' containing only documents taken charge by him and ``Recent File List'' containing his recent viewed documents, which allow users to obtain the documents most important to them and simply continue their respective workflows.

\subsubsection{Data Extraction}\label{data_extraction_model}
Generally \deepshovel~extract data from these parts from literature:
\begin{itemize}
    \item \textbf{Meta Information Extraction} For each uploaded document, \deepshovel~uses multiple parsing tools (e.g., Grobid, Science Parse, and PdfFigures 2.0) to independently extract its meta information and mix all the information with a voting mechanism.
    \item \textbf{Table Extraction:} First, \deepshovel~uses an object detection model Detectron2 \cite{wu2019detectron2} trained on TableBank \cite{li2019tablebank}, a benchmark dataset for table detection, to detect the region of tables. Then for each table, a series of rules are adopted to locate each cell within it. Once users confirm the cell structure of a table, Tesseract \cite{10.5555/1288165.1288167} will be applied to detect the text in each cell and establish the final digitalized table.
    \item \textbf{Text Extraction:} To extract academic entities from papers with the format of PDF, \deepshovel~first utilizes PDFFigures 2.0 \cite{clark2016pdffigures} to parse each text section from the original files. Then some rules and the natural language processing library spaCy \cite{spacy2} are adopted to automatically extract entities of different types from the parsed text sections.
    \item \textbf{Map Recognition and Location Extraction:} Users can box the region of any map they care about. Then \deepshovel~will detect the longitude and latitude labeled at the margin of the map and determine the coordinate range of the entire map. Then if users click any location on the map, the exact coordinates of the location will be automatically calculated and recorded.
\end{itemize}

\section{Evaluation}
We conducted a user study to evaluate \deepshovel. The study examined the following research questions:
\begin{itemize}
    \item \textbf{Q1: Can \deepshovel~cover all the tasks of the data collection from literature?}
    \item \textbf{Q2: Can researchers successfully obtain the data that can build the database?}
    \item \textbf{Q3: Can \deepshovel~help researchers better collaborate in teams?}
\end{itemize}
\subsection{Participants}
We invited 14 users (U1-U14) from 9 different teams (T1-T9) currently using \deepshovel~to join our evaluation. 
They started using \deepshovel~at the same time (according to the registration time).
The demographic characteristics and backgrounds of the participants are reported in Table \ref{tab:evaluation_participants}. 
\begin{table}[]
    \centering
    \begin{tabular}{ccll}
    \toprule
    TID &  UID  & Gender & Role in Team \\
    \midrule
      T1   &  U01 & Male & PhD Student\\
      T1   &  U02 & Female & PhD Student\\
      T1   &  U03 & Male & PhD Student\\
      T2   &  U04 & Female & PhD Student\\
      T2   &  U05 & Male & Associate Professor\\
      T2   &  U06 & Male & PhD Student\\
      T3   &  U07 & Male & Postdoctoral Researcher\\
      T4   &  U08 & Female & PhD Student\\
      T5   &  U09 & Female & PhD Student\\
      T6   &  U10 & Male & PhD Student\\
      T6   &  U11 & Female & PhD Student\\
      T7   &  U12 & Male & PhD Student\\
      T8   &  U13 & Female & PhD Student\\
      T9   &  U14 & Female & Associate Researcher\\
    \bottomrule
    \end{tabular}
    \caption{Demographics of user study participants.}
    \label{tab:evaluation_participants}
\end{table}

\subsection{Procedure}
Each user study session lasted around 45 minutes and was conducted remotely via Tencent Meeting due to the COVID-19 pandemic. 
Participants accessed \deepshovel~using the browser on their computers and shared their screens with the experimenter. 
All sessions are recorded on video.

Considering the participants are all the current users of \deepshovel, the experimenters only briefly introduced the \deepshovel~and the study. 
In order to evaluate the system coverage of extraction tasks, we first asked the participants to extract the data of a representative article selected on their own about their research with \deepshovel. 
As mentioned in \S\ref{Human-AI_Collaboration_for Data_Extraction}, we provided meta information, table, text, and map extractions.
We asked participants to process the file with their usual workflow and observed whether their operating procedures aligned with our system design assumptions.
After the data extraction part, each participant filled out a post-study questionnaire. Then we had a 15-minute semi-structured interview with each group of participants about their team collaboration experience with \deepshovel.

\subsection{Results}
All 9 groups of participants have shown us how they use \deepshovel~to extract data from the scientific literature.
The time used on each participant is about 10 minutes.
There is no apparent deviation between the operation process of each participant and the process we envisioned. 
However, the difference among their research interests means that the data they need to extract is distributed differently in the article, leading to the difference in the function order. 
\textit{Meta Information Extraction} is a function that every participant use. 
Among other functions, most users will use \textit{Table Extraction} first and then \textit{Text Extraction}, while some users will only use \textit{Text Extraction}. 
A small number of users will use the function \textit{Map Extraction}.
Since teamwork is a long process and it is difficult to observe directly, we conducted an assessment in the questionnaire and learned the specific situation in the interview.
We will discuss the results from the post-study questionnaire and the interview in the rest of the section.

\subsubsection{Post-study Questionnaire}
\begin{figure}
    \centering
    \includegraphics[width=\linewidth]{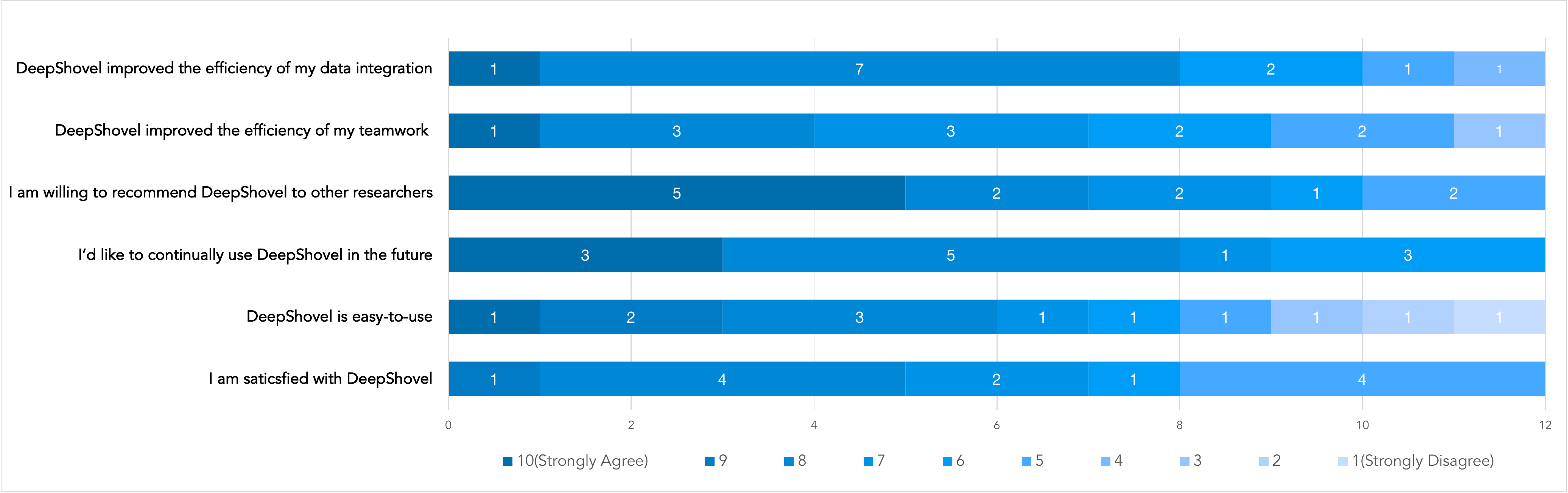}
    \caption{Results of the post-study questionnaire.}
    \label{fig:evaluation_questionnaire}
\end{figure}

We asked all participants to fill out the post-study questionnaire to rate the usability, usefulness, and user experience of teamwork. 
The questionnaire uses a 10-point Likert scale from "strongly disagree" to "strongly agree".
Note that two participants failed to fill out the post-study questionnaire. 
The results of the questionnaire are summarized in Figure \ref{fig:evaluation_questionnaire}. 
Specifically, \deepshovel~scored on average 6.75 (SD=1.42) on "I am satisfied with \deepshovel", 6.33 (SD=2.83) on "\deepshovel~is  easy-to-use", 7.91 (SD=1.44) on "I'd like to continually use \deepshovel~in the future", 8.00 (SD=1.91) on "I am willing to recommend \deepshovel~to other researchers".
And about team collaboration, \deepshovel~scored on average 6.67 (SD=1.75) on "\deepshovel~improved the efficiency of my teamwork", 7.3 (SD=1.59) on "\deepshovel~improved the efficiency of my data integration".
We find that the score of satisfaction is lower than the score of willingness to continue to use, and the overall score is quite different. 
We believe that this is due to different data extraction tasks caused by different research projects (see Table \ref{tab:evaluation_participants}), and we will focus on explaining this problem in the interview results.

\subsubsection{User Experiences and Feedback}
We have a 15-minute semi-structured interview with each group of participants at the end of our user study.
The post-study questionnaire results are discussed, and the questions about user experience (including team collaboration experience) are asked in the interview, which brings some important findings:

\textit{\deepshovel~brings the possibility for geoscientists to form a team to build the database collaboratively.}
Actually, some participants in the user study did not form a team to carry out this work before because of the difficulty of data extraction and the lack of effective team collaboration tools and workflows.
U08 from T4 said "My supervisor only considered forming a data extraction team to advance big data-related research after learning about \deepshovel".
U14 from T9, U12 from T7, and U09 from T5 also mentioned a same situation in their research.
We think this shows that \deepshovel~has helped researchers to promote data-driven research to a certain extent, prompting them to conduct deep-time and deep-space research.
It also proved that our UI design and AI assistance for teamwork are effective.
On the other hand, there are cases of non-team use, such as U07, U04, and U05 from T2 (although they are from the same team, their work does not overlap, and they do not have teamwork tasks).
Even though they do not need the functions for team collaboration, they are also satisfied with the system because it effectively reduces the complexity of data extraction.
No matter it was for team use or personal use, the participants gave a relatively satisfactory score for the AI Assisted Team Collaboration, which is the design that assists in the integration of data.
For example, U01 from T1 said "The integration of data from a single table into a comprehensive table allows me to enter all the data into the database faster".

\textit{Different data requirements lead to different user experience.}
The reason for users' frustration with lower satisfaction in the questionnaire is some technical limitations, such as the parsing problem introduced by PDF files and the accuracy of text extraction.
For example, U07 from T3 mentioned "The data I care about are always in the text, but the region name recognition is very bad". 
U14 from T9 said "The inability to customize tags prevents me from labeling the words I care about". 
Participants were generally satisfied with the table extraction module.
When many participants mentioned the table extraction, they all said "it has greatly improved my work efficiency".
U13 from T8 said "In the past, I thought that completing the data collection work was a distant matter, but now \deepshovel~makes me feel that it is possible and close at hand".
Participants also mentioned that the level of use of different modules affected their satisfaction scores to a certain extent. 
When participants mainly use table extraction to complete tasks, in other words, when the data they care about is mainly in tables, they are more satisfied with \deepshovel.

\textit{Different teams have different collaboration ways when using \deepshovel.}
U01 from T1 said that they have 18 people in the team. They fully used the team management system and basically aligned with our vision.
However, the situations are mainly different when they are in a smaller team.
U10 from T6 said "We did not have distinguished team roles, but collected documents together and divided the tasks equally".
Therefore, they creatively use the take charge function to mark whether the task is complete instead of dispatching the task.
We think this shows that the collaborative design of \deepshovel~can adapt to the use of different team sizes, while also applies to the individual use scenarios as mentioned before.
Furthermore, some functions should be further expanded to mark the completion of the task, which is a function that we do not currently provide.

\textit{Cross-team cooperation took place in the U13 and U14 teams.}
In order to distinguish the groups of participants, we did not indicate this phenomenon in Table \ref{tab:evaluation_participants}.
U13 mentioned that she participated in the research work of three real-life teams, which is common in scientific research.
However, she also mentioned that some of her personal work is also stored in \deepshovel~for convenience.
She said "This actually caused confusion in file management to some extent".
We think this is also an interesting phenomenon that we never imagined.
In the future, we should consider improving the convenience of cross-team collaboration to avoid this kind of file management confusion.

\section{Discussion}
The results from our user study and the practical application suggest that geoscientists can successfully collaborate with \deepshovel~to extract data from scientific literature and integrate the data into databases. \deepshovel~also performed well in helping researchers' cooperation in their team. 
In this section, we discuss the lessons we learned and the design implications of our work.

\subsection{Data Extraction from Scientific Literature with Human-AI Collaboration}
\subsubsection{Why End-to-End is Not a Good Choice?}
Due to the accumulation of errors in end-to-end approaches, the final outcome might be unacceptable.
Meanwhile, manually checking and correcting these errors is a very tedious and difficult job.
We would like to claim that we do not believe today's AI technology can build a "fully automated" system to replace researchers in data extraction.
To ensure the quality of the database used in further research, researchers still have to clean and correct data manually.
We think that building a human-AI collaboration solution with the appropriate level of automation would be a better way to solve the problem so that the human and AI can jointly iterate, improve and complete the data extraction.
The user makes the final decision of all data extraction, and AI fully follows the user's instructions in this interaction process to ensure the accuracy of the data.
For example, in the table extraction process, AI only recommends where the tables are, and users decide which table they would like to extract.

\subsubsection{Decision Making in Human-AI Collaboration}
\begin{figure}
    \centering
    \includegraphics[width=\linewidth]{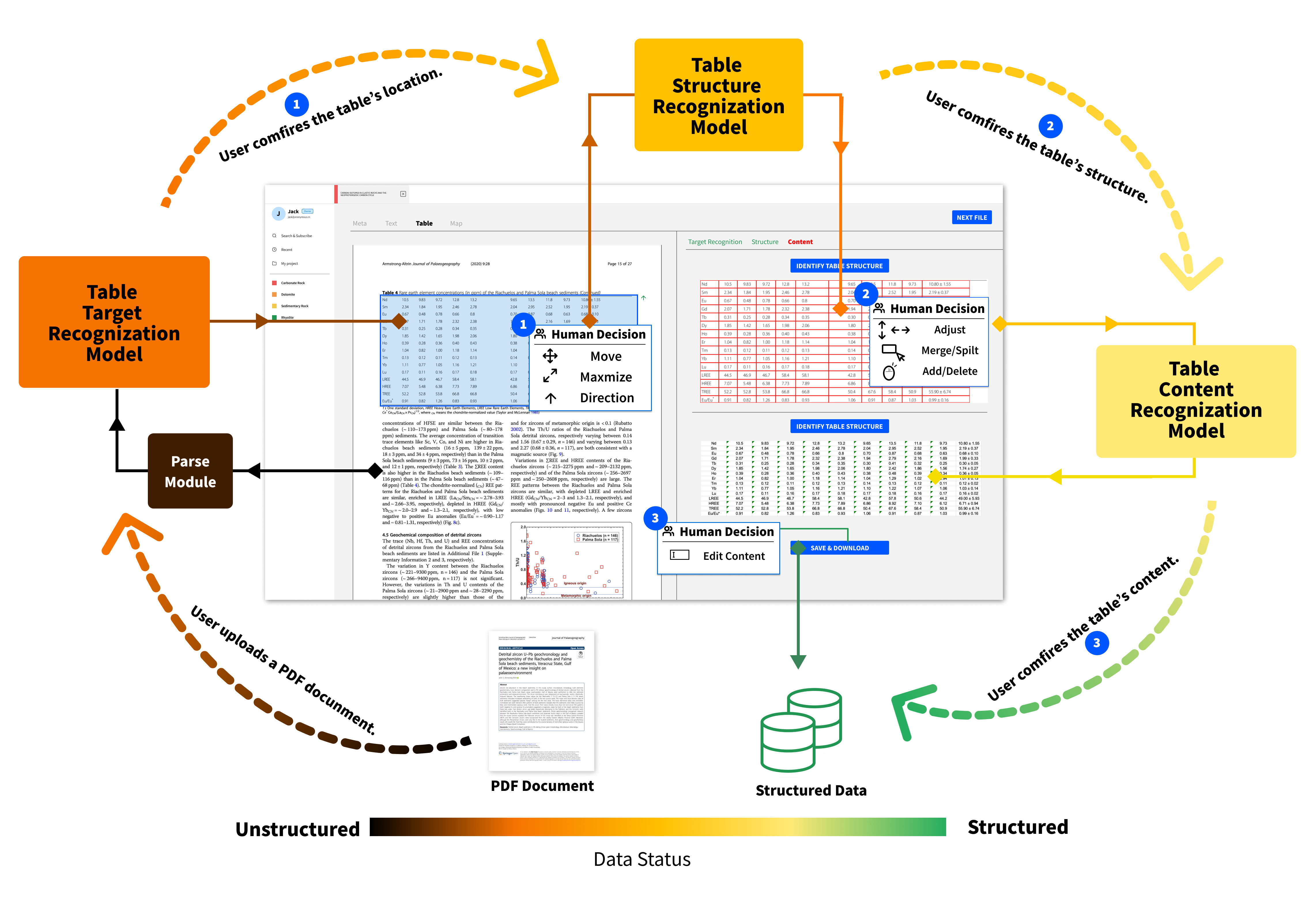}
    \caption{The step-by-step interaction in table extraction.}
    \label{fig:table_extraction_interaction}
\end{figure}

In \deepshovel, AI only suggests the tables' structure and content, and users can edit this information and decide on the final output.
The user will take turns interacting with different models during a table extraction, each model explaining its own task to the user.
Such an interactive form can first effectively help users understand the role of the model, allowing users to make decisions more intuitively as to whether they need to be modified.
Secondly, due to the step-by-step interaction (Figure \ref{fig:table_extraction_interaction}), information overload is avoided, and the user can focus more on the current decision.
Considering the interaction process between models and humans, preventing information overload and model overload, and making the interaction process as disassembled as possible to fit into the human decision-making process are two critical issues to be considered.

\subsection{Team Collaboration and Task Distribution}

\subsubsection{Fine-grained Functional Division}
There may be many different forms of task assignment in multi-player cooperative tasks. If the framework design restricts task division, it will cause damage to the original team operation mode.
When each user performs a data extraction task in a single PDF document, \deepshovel~follows the rules of the distribution of target data in the document and breaks down the task into meta information, text, table, and map extraction, and each subtask is served by a different model.
The models are introduced by task division so that users do not interact with multiple models at once.
Such a design can avoid users from processing information from three different modalities: image, text, and table at one time. 
This splitting of functional modules based on data distribution brings greater flexibility to team collaboration while reducing the pressure on online services.
In \deepshovel, the team can divide labor according to projects, documents, or even tasks, which adapts to the existing team cooperation methods to the greatest extent.
We believe that when designing related collaboration platforms in the future, different forms of team collaboration should be considered in the process design.

\subsubsection{AI Assisted Team Collaboration}
Our research found that no tool exists in current team data integration, and geoscientists have to integrate data using Excel manually.
We think this is an essential step after data extraction, but there is still no very reasonable solution for some researchers who are just average computer users.
However, there are still problems with this design.
Some users report that data cannot be merged correctly because different authors have different descriptions of some fields when drawing tables, and the information in the text cannot be merged.

\subsubsection{Information Sharing in In-team Collaboration}
In team collaboration tasks, information sharing within the team is crucial to the efficiency and quality of task completion.
We studied the teamwork of researchers in the task of data extraction and found that they were in a relatively primitive state in the way they shared raw data and managed files.
We believe that online platforms are very necessary in the current situation affected by the COVID-19 pandemic.
The online system also allows researchers to work without geographical constraints.
\deepshovel~uses shared projects and files online to maximize the exposure of team internal data to all team members.
Under the premise of this openness, we use the file lock mechanism and the file principal mechanism to ensure the consistency of operations during data extraction.

\subsubsection{Cross-team Collaboration Experience}
We also studied the case of cross-team collaboration and supported this collaboration in \deepshovel.
Our design of user personas gives team leaders control over different members. 
In the case of cross-team collaboration, the team leader can control the permissions of team members and the security of data to the greatest extent, and users can also establish a new team within the system to support cross-team cooperation.
However, we still have users reporting that they want their data to remain private until the end of the study, so we put some restrictions on collaborating with cross-teams. 
With the emphasis on open science \cite{10.1145/3462204.3481785}, this is still an issue that needs to be considered.
When dealing with cross-team collaboration, how to balance the sharing of information with the confidentiality of scientific research can be an important issue.

\section{Limitations and Future Work}

Our user research and system function design are both conducted in the field of geoscience, especially in the \DDE~program.
We propose a novel and general collaborative framework for scientific literature data extraction in natural science.
However, when extended into other disciplines, there may be more problems and difficulties that we still do not fully understand, including the specific data extraction tasks and the form of team collaboration.
We plan to expand \deepshovel~to other disciplines in the future as more extensive user research is required.

Besides, since our research is conducted remotely through an online meeting service, the setting of such scene may be quite different from the actual situation.
We may lack observation of the working status of their team cooperation.
For example, we cannot observe how the team communicates in their actual state, how the data is transmitted throughout the teamwork process, and how their data integration process is actually accomplished.
Moreover, We do not conduct a comparative experiment study by comparing \deepshovel~with some baselines, including extracting data manually and using existing tools like ABBYY.
Since the effect of teamwork and the efficiency of data extraction requires long-term observation to obtain, we think we need to observe the team's feedback after long-term use and conduct a larger-scale field deployment.

The rest are some technical limitations in \deepshovel:
1) the quality of data extraction is greatly affected by the quality of PDF files, and we cannot handle some low-resolution scans that are too old;
2) most AI models in \deepshovel~are based on rules provided by geoscientists and a relatively small amount of geoscience data, which may lead to some problems in the processing of uncovered literature;
3) in the current proof-of-concept stage, \deepshovel~does not meet all the types of data demands (e.g., points location in some scatterplots) because of the lack of relevant datasets.
Fortunately, such problems exist in different independent data extraction modules, not affecting the system design framework.
In the future, we will continuously add new modules and improve existing modules through rapid system iterative upgrades.

\section{Conclusion }
In this paper, we present \deepshovel, an online collaborative platform for data extraction in the scientific literature with AI assistance that can help researchers cooperate with their teammates to extract data from PDF documents and build a scientific database.
The design of \deepshovel~is motivated by the user research we place in the field of geoscience.
\deepshovel~can help researchers extract and aggregate data containing meta information, tables, texts, and location from the literature.
The research team can collaborate in \deepshovel, and team members can share resources and progress with others.
\deepshovel~has been deployed for one month and there are already 253 users from 36 geoscientist teams within the \DDE~program use it in a daily basis.
More than 240 projects and 46,000 documents are being processed for building scientific databases.
The follow-up user evaluation with 14 researchers confirms that \deepshovel~improves researchers' efficiency in data extraction from geoscience literature and promotes the close collaboration of their teams.

\begin{acks}
We owe a particular debt of gratitude to the scientists from the Deep-time Digital Earth project who all contributed enormously valuable feedback. 
We also thank Jia Guo, Yifei Shen, Qi Li, Zhixin Guo, Mingxuan Yan, Mingze Li, Le Zhou, Jingyao Tang, Han Liu, Shengling Zhu and Tao Shi for their support to our system development.
This work is supported by National Natural Science Foundation of China (No.42050105, No.62106141) and Shanghai Sailing Program (21YF1421900).
\end{acks}

\bibliographystyle{ACM-Reference-Format}
\bibliography{deepshovel}



\end{document}